\documentclass{statsoc}
\pdfoutput=1

\usepackage[a4paper]{geometry} % required for statsoc+pdflatex
\usepackage{vmargin} % required for statsoc+pdflatex
\RequirePackage[T1]{fontenc}
\RequirePackage{amsmath,amssymb}
\RequirePackage{natbib}
\RequirePackage[colorlinks,citecolor=blue,urlcolor=blue]{hyperref}

\usepackage{graphicx,epsfig}
\usepackage[font=footnotesize,textfont={footnotesize,sl},labelfont=bf]{caption}
\usepackage{subcaption}

\numberwithin{equation}{section}
\newtheorem{thm}{Theorem}[section]
\newtheorem{defn}[thm]{Definition} 
\title[A procedure to detect general association]{A procedure to detect general association based on concentration of ranks}
\author[P. Rudra and F.A. Wright]{Pratyaydipta Rudra}
\address{Department of Biostatistics, University of North Carolina at Chapel Hill}
\author[P. Rudra and F.A. Wright]{Fred A. Wright \footnote{{\it Address for correspondence:} Fred A. Wright, NCSU Bioinformatics Research Center, 2601 Stinson Drive, Campus Box 7566, Raleigh, NC 27695. E-mail: fred\textunderscore wright@ncsu.edu}} 
\address{North Carolina State University Bioinformatics Research Center}
\begin{document}
\maketitle

\begin{abstract}
\noindent In modern high-throughput applications, it is important to identify pairwise associations between variables, and desirable to use methods that are powerful and sensitive to a variety of association relationships. We describe RankCover, a new non-parametric association test for association between two variables that measures the concentration of paired ranked points.  Here `concentration' is quantified using a disk-covering statistic that is similar to those employed in spatial data analysis. 
%RankCover  valid for any form of association and does not depend on the marginal distribution of the variables. 
Analysis of simulated datasets demonstrates that the method is robust and often powerful  in comparison to competing general association tests. We illustrate RankCover in the analysis of several real datasets. 
%A method of testing the association of two variables while controlling the effect of a third variable is developed using RankCover. 
%We have illustrated how this procedure controls the type-I error and achieves reasonable power at the same time. 
\end{abstract}

\keywords{General Association; Rank-based Test; Test of Independence}

%%% Introduction %%%

\section{Introduction}
\label{sec:intro}

The need for statistical methods to identify general pairwise association is increasingly recognized, as evidenced by recent attention to methods such as distance correlation (dCor) \citep*{szekely2007,szekely2009}, Maximal Information Coefficient (MIC) \citep*{reshef2011}, and the Heller-Heller-Gorfine (HHG) method \citep*{heller2013}.  The term {\it general association} refers to any departure from independence among random variables, and methods differ in the types of departures to which they are sensitive. The need for general association tests is perhaps greatest for analysis of large datasets, for which discovery-based approaches are needed, without prior hypotheses regarding the form or structure of dependence.  In addition to the need to test dependence among pairs of variables as a primary analysis,  dependencies can invalidate inference for downstream methods that require independence among input variables \citep*{albert2001}. 

%Classical parametric methods, such as Wilk's lambda \citep*{wilks1935} or the Puri-Sen likelihood ratio test %\citep*{puri1971} suffers from the disadvantage of requirement of distributional assumptions. On the other %hand, classical nonparametric methods are either not robust to wide range of alternatives or have very low %power. 

Standard parametric and non-parametric tests of association, such as linear trend testing \citep*{mann1945,kendall1975,cuzick1985,hamed1998}, are sensitive to only specific alternatives, while classical tests of association
(\citep*{wilks1935} \citep*{puri1971} are not distribution free.  Recent work has tried to capture the measure of association through a generalized correlation coefficient, which is able to capture numerous forms of relationships. MIC \citep*{reshef2011} and dCor \citep*{szekely2007,szekely2009} are two recently introduced measures of general association. HHG \citep*{heller2013}  has been shown to be powerful against many alternatives, and also shown to be consistent, in the sense of having increasing power with sample size $n$, against all dependent alternatives. However its performance in small to moderate samples against varying alternatives has not been studied.

\begin{figure}[!ht]
\centering
\begin{subfigure}[b]{\textwidth}
                \centering
                \includegraphics[width=\textwidth]{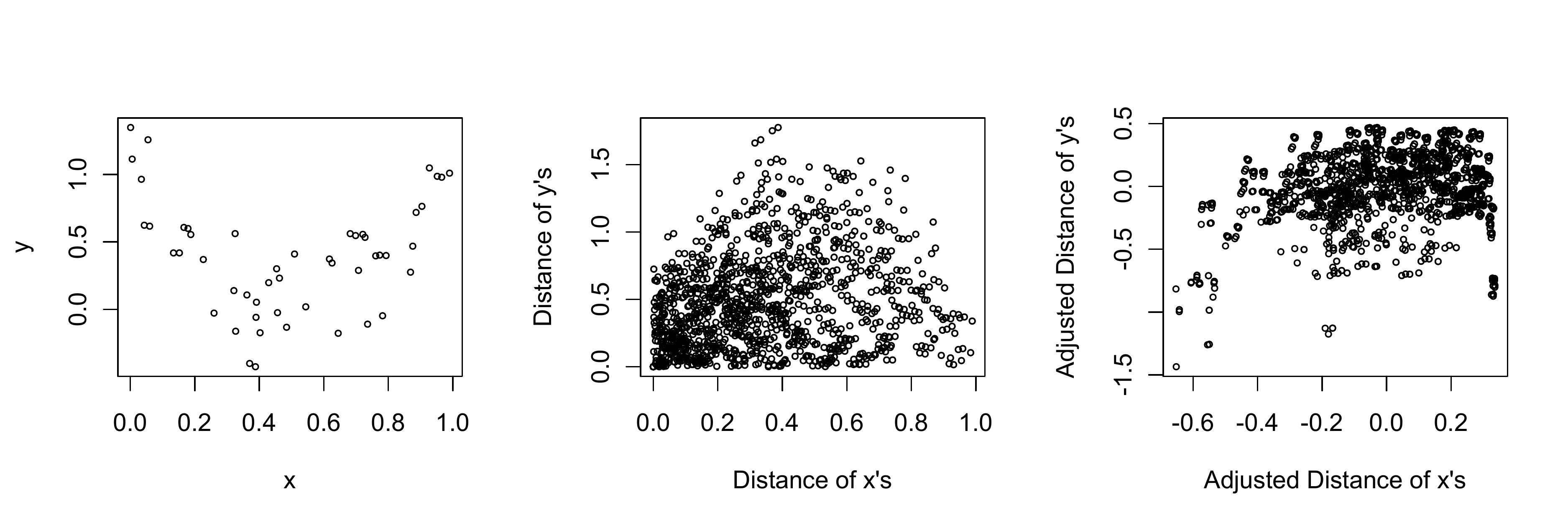}
                %\label{fig:1,1}
        \end{subfigure}\\
\begin{subfigure}[b]{\textwidth}
                \centering
                \includegraphics[width=\textwidth]{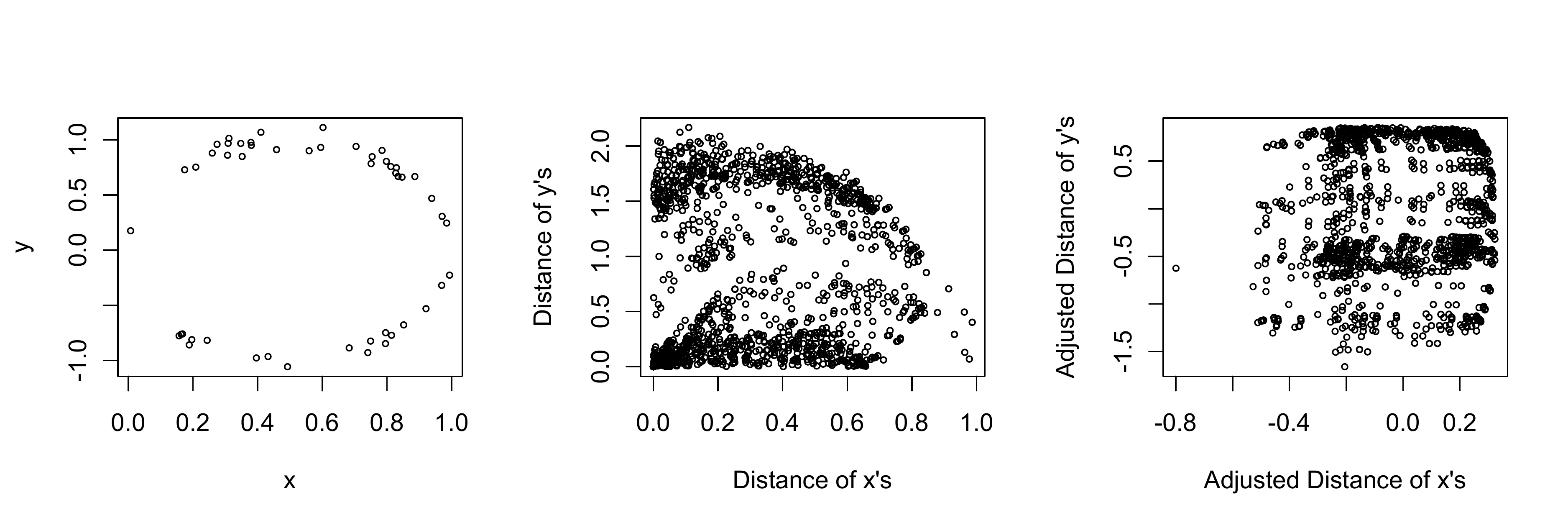}
                %\label{fig:1,2}
				\end{subfigure}\\
        
\caption{Illustration of paired adjusted distances underlying dCor. (top row) Illustration of dCor for a quadratic relationship between x and y.(bottom row) A circular relationship between x and y. The adjusted paired distances show little correlation.}
\label{fig:1}
\end{figure}

In providing motivation for MIC,
\citet*{reshef2011} showed that it is {\it equitable} in the sense that the value of the coefficient is similar for various forms of association that are equally ``noisy" in their departure from a functional relationship.
However, \citet*{simon2014} argued that such equitability makes the procedure less powerful in a number of situations, and that the distance correlation (dCor) is in fact more powerful in almost all of a variety of simulated associations that they studied. 
%Based on their note, it is clear that for the hypothesis testing problem, MIC is inferior as compared dCor in %most situations.
A potential weakness of dCor is that it is not powerful to detect nonmonotone relationships, such as a circle \citep*{santos2013}, and the detection of such relationships is a primary motivation to develop methods for general association.
To gain insight into why dCor loses power in case of nonmonotone alternatives, it is helpful to consider dCor from a geometric point of view. dCor is motivated by consideration of distances of the empirical characteristic function under the null vs. under the alternative.  For observed data, the dCor statistic is the Pearson correlation of distances  (after some adjustments) between all pairs of samples. For an observed random sample $(x,y)={(x_k,y_k):k=1,2,...,n}$, 
%X \in \mathbb{R}^p$ and $Y \in \mathbb{R}^q$, 
the distances between pairs of samples are defined as $a_{kl}=|x_k-x_l|$ and $b_{kl}=|y_k-y_l|$; $k,l=1,2,...,n.$ 
%For the one dimensional case, a correlation of such distances $a_{kl}$ and $b_{kl}$ is intuitively appleaing since under the alternative, the %pairs which are close in the direction of $X$ axis are expected to be close in the direction of $Y$ axis also. 
%
The approach is intuitively sensible when the relationship is monotone, as sample pairs that are close on the $x$-axis should also be close on the $y$-axis.  However, for non-monotone relationships, pairs of points that are close on the $x$-axis can be quite distant on the $y$-axis (\autoref{fig:1}).
%[PRATYAY: WHY DO WE EVEN MENTION p AND q? JUST TO MAKE THINGS LOOK MORE GENERAL? IF WE NEVER DO THIS IN HIGHER DIMENSIONS WE SHOULD JUST DROP THAT NOTATION]

% The adjustment somewhat takes care of the nonmonotonicity (\autoref{fig:1,1}), but still remains unsuccessful in some cases %(\autoref{fig:1,2}).  

%The nature of the measure dCor is global in the sense that it depends on the exact distance of all pairs of %observations. 

Another way to approach the general association problem is to consider spatial randomness of
points $(x_k, y_k)$, and the proposed tests of general association attempt to be sensitive to alternatives in which the points are clustered or otherwise closer to each other than expected under no association.
A class of testing procedures sensitive to local clustering has been devised in the field of spatial statistics \citep*{clark1954, holgate1965, holgate1965new, ripley1979, smith2004}. The $F$ function by Diggle \citep*{diggle1983} uses nearest neighbor distances to devise a test against the hypothesis of complete spatial randomness. 
%
%Even though our null hypothesis is not exactly same as the hypothesis of complete spatial randomness, %these ideas can be applied in our case as well. Diggle proposed Monte Carlo methods of testing with the %$G$ and $F$ functions, the distribution functions of nearest neighbor distances and point to nearest event %distances respectively. Such testing procedures by comparing the empirical distribution function to the %theoretical one using permutation tests is computationally demanding and their performance in terms of %power and robustness is unclear. A method devised by Clark and Evans \citep*{clark1954} based on mean %nearest neighbor distances along with a large sample approximation \citep*{donnelly1978} saves the %computation cost. However the large sample approximation lies on the assumption that the variables are %uniformly distributed under the null.
Adapting these ideas from spatial analysis, we propose RankCover, a method that quantifies the concentration of $(x,y)$ values by measuring the area covered by laying disks of a fixed radius over each point in the scatter plot of the ranks of the two variables.
We demonstrate that RankCover is robust in the sense that it has power against a variety of alternatives, and  regardless of the marginal distributions of the two variables.  Moreover, RankCover has favorable power in comparison to dCor, MIC and HHG, and is especially useful for detecting oscillating relationships.  Simulations demonstrate that RankCover and dCor are in some sense complementary, and a hybrid of the two methods is a robust  procedure that is powerful for most association types of interest.

% In \autoref{sec:partial} we have proposed a method to test the association of two variables after adjusting %for a possible confounder. The method works for all forms of associations between the three variables. The %performance of this method is illustrated using simulated data.

\section{Methods}
\label{sec:methods}

\subsection{Motivation}
\label{sec:motivation}

%While testing the association of two variables, thinking about the problem geometrically makes much sense. %A scatter plot of the ranks of the two variables is likely to show some clustering when they are associated as %compared to when they are independent. One should note that this idea does not hold if the rank scale is %not considered. The other extreme, where one observation in the scatter plot inhibits the presence of %another observation near it, might be an interesting alternative for the spatial statisticians. For our %applications, we can ignore such alternatives and consider testing the one sided alternative. Even though we %are thinking along the line of complete spatial randomness, our set up differs from that of testing complete %spatial randomness in two aspects. First, we are interested in only one sided alternatives which is known as %aggregation in the spatial statistics literature. Also, the complete spatial randomness assumes that the %points follow a Poission point process or equivalently they are uniformly distributed in a region. This is not %true in general for the null hypothesis of our interest. However, using the test statistics to test complete %spatial randomness makes sense in the general association testing problem if they are applied on the ranks %of the two variables. 

RankCover starts by computing ranks of the original $x$ and $y$ values, and we assume there are no tied values. The use of ranks considerably simplifies the problem, by placing the intervals between successive ranked values on a common scale.  In addition, for ranked values, the null distribution depends only on the sample size $n$. Thus the only computation lies in computing the observed statistic, while the null distribution can be pre-computed and is applicable to any dataset of size $n$. 

\begin{figure}[!ht] 
                \centering
                \includegraphics[width=0.7\textwidth]{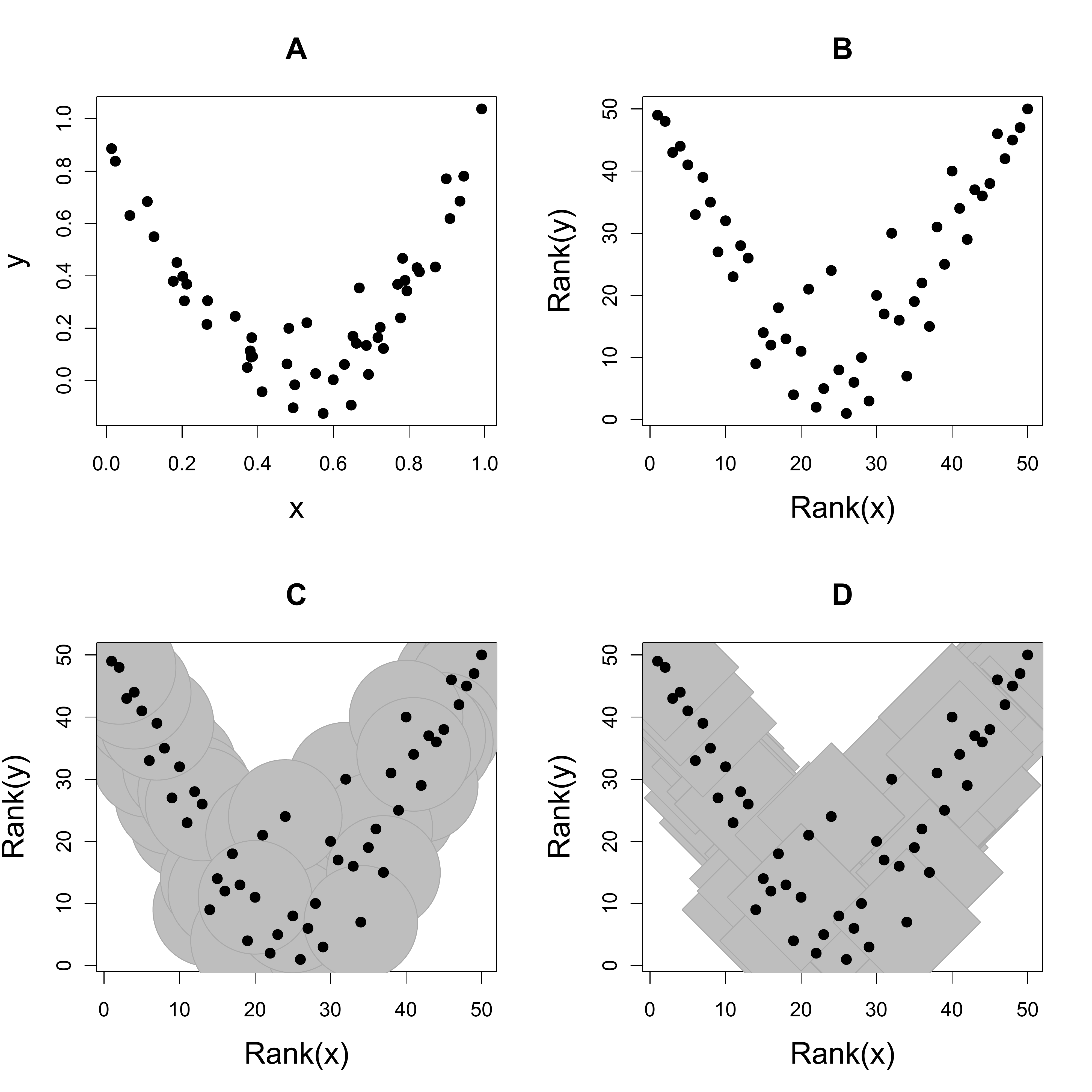} 
                \caption{Illustration of RankCover for sample size $n=50$: \textbf{A.} Scatter plot of the two variables. \textbf{B.} Scatter plot on the rank scale \textbf{C.} Disks laid on the scatter plot on rank scale using Euclidian distance \textbf{D.} Disks laid on the scatter plot on rank scale using Manhattan distance.}     
                \label{fig:1prime} 
\end{figure}

Diggle's $F(\delta)$ function as introduced in \citep*{diggle1983} is the distribution function of the distance between a randomly chosen point in a region to the nearest observed point $(x_k,y_k)$.
To obtain an empirical estimate of the $F(\delta)$, the investigator conceptually lays disks of radius $\delta$ on each point $(x_k, y_k)$ and calculates the proportion of the surrounding region covered by the union of the disks (\autoref{fig:1prime}).  If $x$ and $y$ are highly associated, the areas covered by the disks should be small, and therefore RankCover rejects only in the left tail of the statistic described below.

Different distance metrics can be used for this purpose and the shape of the disks depend on the choice of the distance metric. For instance, Euclidean distance leads to circular equidistance contours, resulting in circular disks, while the disks are diamond-shaped for Manhattan distance (\autoref{fig:1prime}).

\subsection{The test statistic}
\label{sec:teststat}
The empirical estimate of $F$ can be obtained using the proportion of area covered by the disks.
Acknowledging the discrete nature of the ranks, we consider only the $n\times n$ grid of possible
rank pairs, $\{1,2,...,n \} \times \{1,2,...,n \}$, and whether each of these values on the grid is covered by at least one disk.
Let $(x_{k},y_{k})$ denote the ranks of the $k$th sample pair, $k=1,2,...,n$.

\begin{defn}
Define $d(i,j,x_k,y_k)$ = distance between the point $(i,j)$ on the grid and $(x_k,y_k)$; $d_{ij}=\min_{k} d(i,j,x_k,y_k)$
\end{defn}

\noindent Using this definition, a reasonable statistic for fixed $\delta$ is 
$$\hat{F}(\delta) = \frac{1}{n^2}\sum_{i=1}^n \sum_{j=1}^n {I(d_{ij} \leq \delta)},$$
\noindent where $I(.)$ is the indicator function.

The choice of disk size $\delta$ is an important consideration which has not been fully addressed in the spatial statistics literature. \citet*{diggle1983} suggested computing the entire empirical curve $\hat{F}(\delta)$ to develop a new summary statistic to compare against the null curve. However, this approach makes the procedure prohibitively computationally expensive,
and we propose (See Supplementary Article for details) using a fixed $\delta=\sqrt{n}$ for Euclidean distance (\autoref{sec:parameter}), 
with slight modification under Manhattan distance.  In addition, we modify the statistic to account
for edge effects of the grid, using an $(n+\lceil \delta \rceil)\times (n+\lceil \delta \rceil)$ grid extending beyond the range of the scatterplot. Here $\lceil \delta \rceil$ is the smallest integer greater than or equal to $\delta$. 
Finally, our modified test statistic is
\begin{equation}
\label{eq:2}
T(\delta) = \frac{1}{n^2}\sum_{i=1-\lceil \delta \rceil}^{n+\lceil \delta \rceil} \sum_{j=1-\lceil \delta \rceil}^{n+\lceil \delta \rceil} {I(d_{ij} \leq \delta)},
\end{equation}
where the range of $\{i,j\}$ reflects the outer boundaries of a larger region to account for edge effects.
The null distribution of $T$ depends entirely on $n$, so tables based on simulated null distibutions can be precomputed for various sample sizes (See Supplementary Article).
% Therefore one can simply look into the tables instead of using permutations which saves a lot of computation time. These advantages outweigh the loss of information due to use of ranks.

%In the next section, we will discuss how it helps to choose the disk radius $\delta$ as well 

\subsection{Choice of parameters and distance metric}
\label{sec:parameter}
For the distance metric $d$, we consider here both Euclidian and Manhattan distances, for which later simulations show similar performance (See Supplementary Article). However, the Manhattan distance has advantages in approximating tail areas (See Supplementary Article). Therefore we recommend its use and here present results using Manhattan distance.

\section{Results}
\label{sec:results}

We applied the RankCover method using the Manhattan metric to simulated and real datasets, following setups similar to \citet*{simon2014}, investigating dCor, HHG, and MIC as competing approaches. 
%We have used Manhattan distance for analyzing both simulated data and real data. 
The simulation results indicate that RankCover and dCor have some complementary characteristics, and so we additionally propose a hybrid statistic using
results from  RankCover and dCor. The hybrid method uses
the minimum $p$-value from RankCover and rank-based dCor as a new statistic. 
In addition to simulated data, we illustrate all the approaches on several real datasets.

\begin{figure}[!ht]
                \centering
                \includegraphics[width=0.8\textwidth]{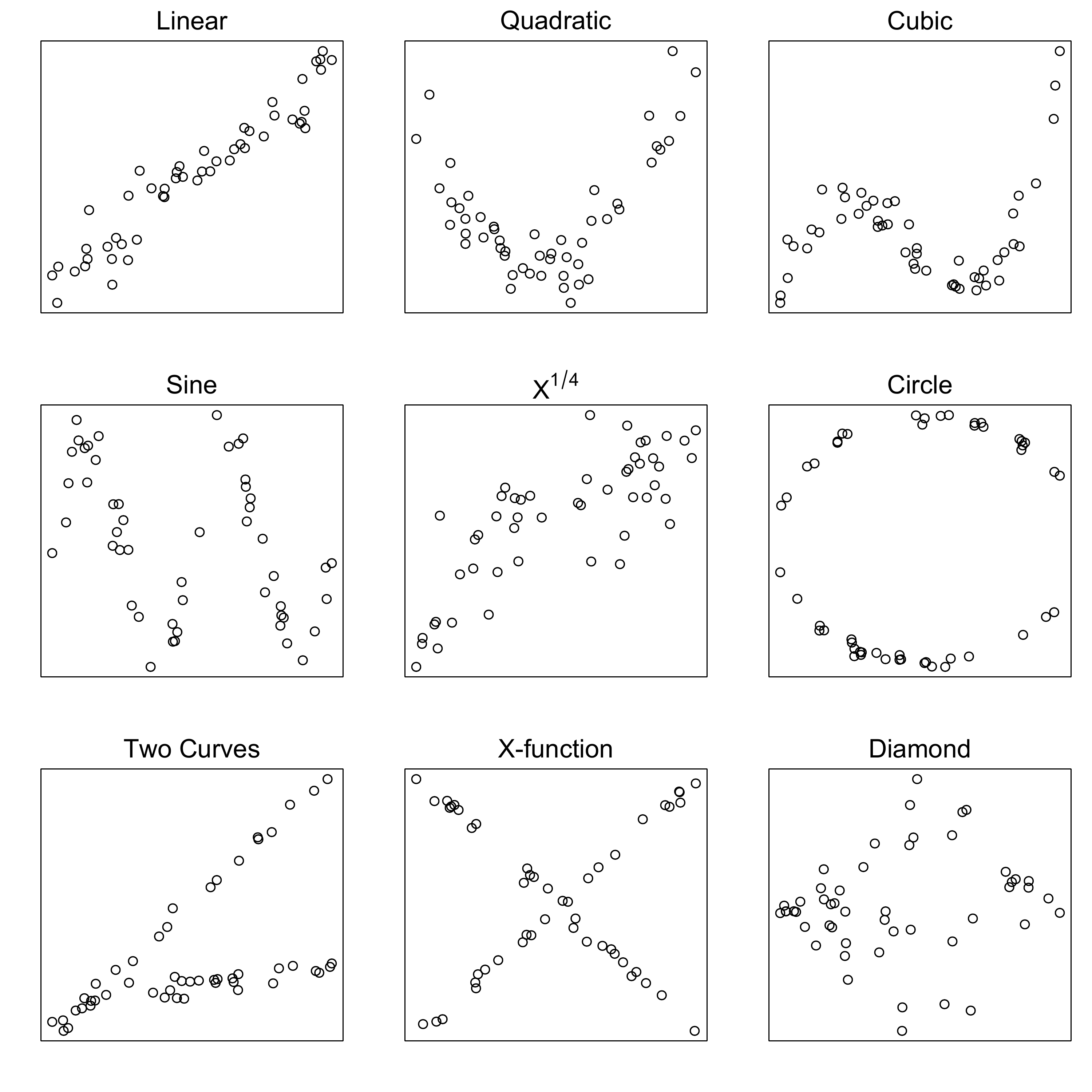}
                \caption{Showing the scatter plots for different relationships between the pair of variables (low noise level)} 
                \label{fig:3} 
                       
\end{figure}

\subsection{Simulation results}  
\label{sec:simulation}
Following the simulation procedure used in \citet*{simon2014}, we have simulated pairs of variables with several canonical dependency relationships (\autoref{fig:3}) and with varying noise levels. In each scenario, the $X$ values were simulated iid from a uniform distribution, while the noise distribution was Gaussian. However, the overall results were similar for
other distributional forms (See Supplementary Article). 
%[PRATYAY: WHAT IS ALPHA? CAN YOU SAY HERE SOMETHING ABOUT ON WHAT SCALE "NOISE LEVEL" IS MEASURED?]

\begin{figure}[!ht]
                \centering
                \includegraphics[width=\textwidth]{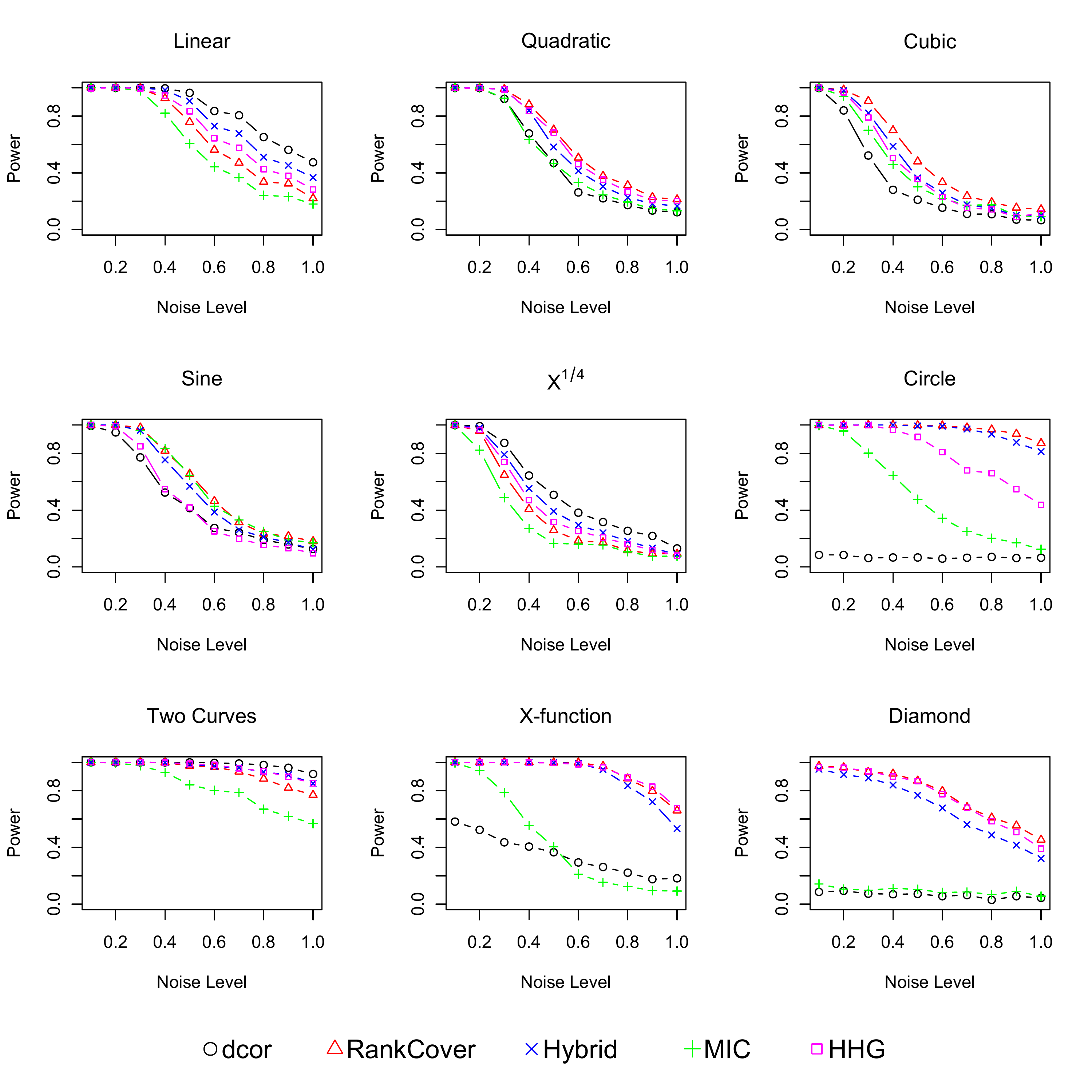}
                \caption{Showing the power of different methods (type-I $\alpha=0.05$ )against different relationships at varying noise levels.} 
%[PRATYAY, REORDER THE METHODS. IT DOESN'T MAKE SENSE TO HAVE MIC SPLITTING UP THE OTHER METHODS
                \label{fig:4} 
                       
\end{figure}

\autoref{fig:4} shows the power for the methods for various relationships, with varying noise levels, for sample size $n= 50$. Here the ``noise level" is a scale quantity appropriate to each relationship form, following \citet*{simon2014} (See Supplementary Article). It is evident that RankCover performs better than MIC in all the situations we have considered. It is found to be more powerful than dCor and HHG in several cases while these methods are found to be more powerful in other cases. Even when dCor or HHG is more powerful, RankCover still has reasonable power to identify the association.
Numerous illustrations provided in the supplementary article indicate that these observations hold true for varying sample sizes, levels of noise, and functional forms for the originating $X$ and noise distributions.

A  careful look into the results indicate that dCor is more powerful than RankCover when the type of association is monotone. When the relationship is non-monotone, dCor is typically not as powerful. We attribute this behavior to the fact that dCor is less sensitive to non-monotone relationships for the reasons described earlier. We have also shown that with monotone relationships the Spearman's rank correlation is as powerful as dCor (See Supplementary Article). Therefore, one might simply use Spearman's rank correlation if there is prior knowledge that the relationship is monotone. 
 On the other hand, RankCover is more sensitive to local clustering of points rather than trends. Thus, it is powerful against even non-monotone relationships like cubic, circular or the ``X'' relationship. 

These observations motivate the use of a hybrid method utilizing both RankCover and dCor, as the two methods appear powerful in different situations. Formally,
a new statistic is defined $s_{hybrid}={\rm min}(p_{dCor},p_{RankCover})$, where $p_{RankCover}$ is the $p$-value obtained by using RankCover, and $p_{dCor}$ is that using dCor on $(rank(x),rank(y))$. The $p$-value for the hybrid method is $p_{hybrid}=P(S_{hybrid}\le s_{hybrid})$. As with RankCover, the $p$-value can be obtained by using pre-computed simulations. 
The hybrid method, as expected, is always less powerful than the most powerful statistic for each scenario, but seems to be robust against all forms of association investigated.

The HHG method also appears to be relatively robust.  However, the ability of RankCover and the hybrid method to detect periodic relationships and non-functional relationships makes it very useful against such alternatives. The fact that RankCover is especially  powerful against periodic relationships will be reinforced by the results in \autoref{sec:enso} and \autoref{sec:yeast}.

We summarize by emphasizing that RankCover and the hybrid method are powerful and robust in comparison to competing methods, and that these simulations cover a large range of relationships and noise levels. The broad conclusions are also not very sensitive to the marginal distributions of $X$ and the error distributions (See Supplementary Article).

%PRATYAY, HERE SAY SOMETHING ABOUT THE RESULTS SHOWN IN THE SUPPLEMENT. EVEN IF THEY JUST REINFORCE THE RESULTS SHOWN IN THE MAIN MS.

\subsection{Real data}  
\label{sec:realdata}

\subsubsection{Example 1: Eckerle4 data}
\label{sec:eckerle4}

We show data from a study of circular interference transmittance \citep*{eckerle1979} from the NIST Statistical Reference Datasets for non-linear regression. The data were analyzed by \citet*{szekely2009} to illustrate dCor, and contain 35 observations on the predictor variable wavelength and the response variable transmittance.  
%[PRATYAY, I FORGOT WHY YOU CHOSE THIS DATASET]

\begin{figure}[!ht]
                \centering
                \includegraphics[width=7cm]{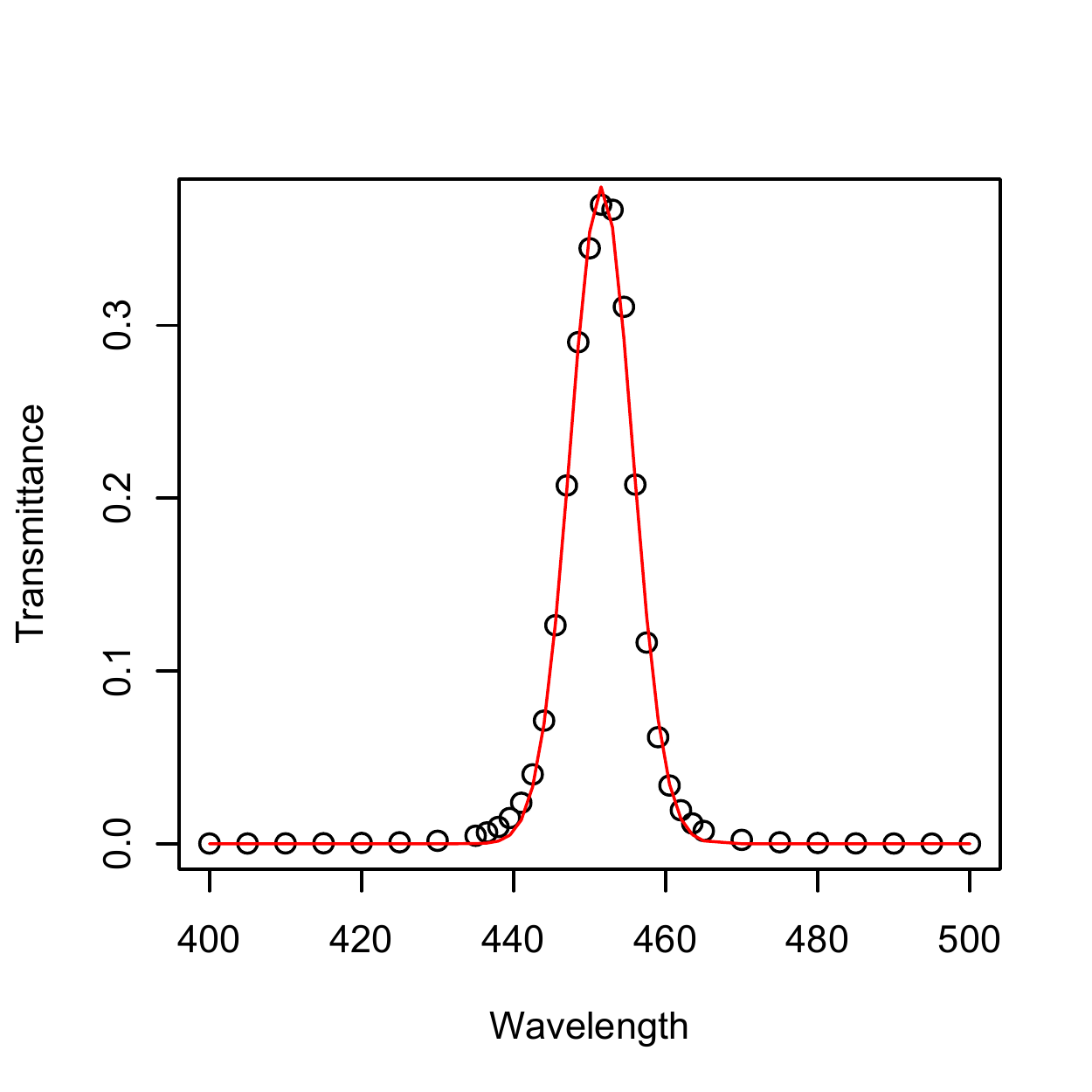}
                \caption{Showing the scatter plot and the fitted curve for the Eckerle4 dataset} 
                \label{fig:6} 
                       
\end{figure}

\autoref{fig:6} shows the scatter plot of the predictor and the response along with the fitted curve (\href{http://www.itl.nist.gov/div898/strd/nls/data/eckerle4.shtml}{NIST StRD for non-linear regression}) based on the model

\begin{center}
$y=\frac{\beta_1}{\beta_2} exp\{\frac{(x-\beta_3)^2}{2\beta_2^2}\}+\epsilon$,
\end{center}
 
where $\beta_1,\beta_2>0, \beta_3 \in \mathbb{R}$ and $\epsilon$ is random Gaussian noise. 
%[WAS THE NOISE ASSUMED NORMAL?].

From the plot, it is evident that there is a very strong non-linear relationship between the two variables.  For dCor, $p= 0.02072$, while  MIC and HHG have $p$-values $<10^{-5}$. The RankCover method and the hybrid method are also highly significant, with $p<10^{-5}$.  
%[HOW DID YOU GET SUCH A SMALL PVALUE WITH ONLY $10^5$ PERMUTATIONS?]

\subsubsection{Example 2: Aircraft data} %[PRATYAY: PUT THIS EXAMPLE SECOND]
\label{sec:aircraft}

\begin{figure}[!ht]
                \centering
                \includegraphics[width=\textwidth]{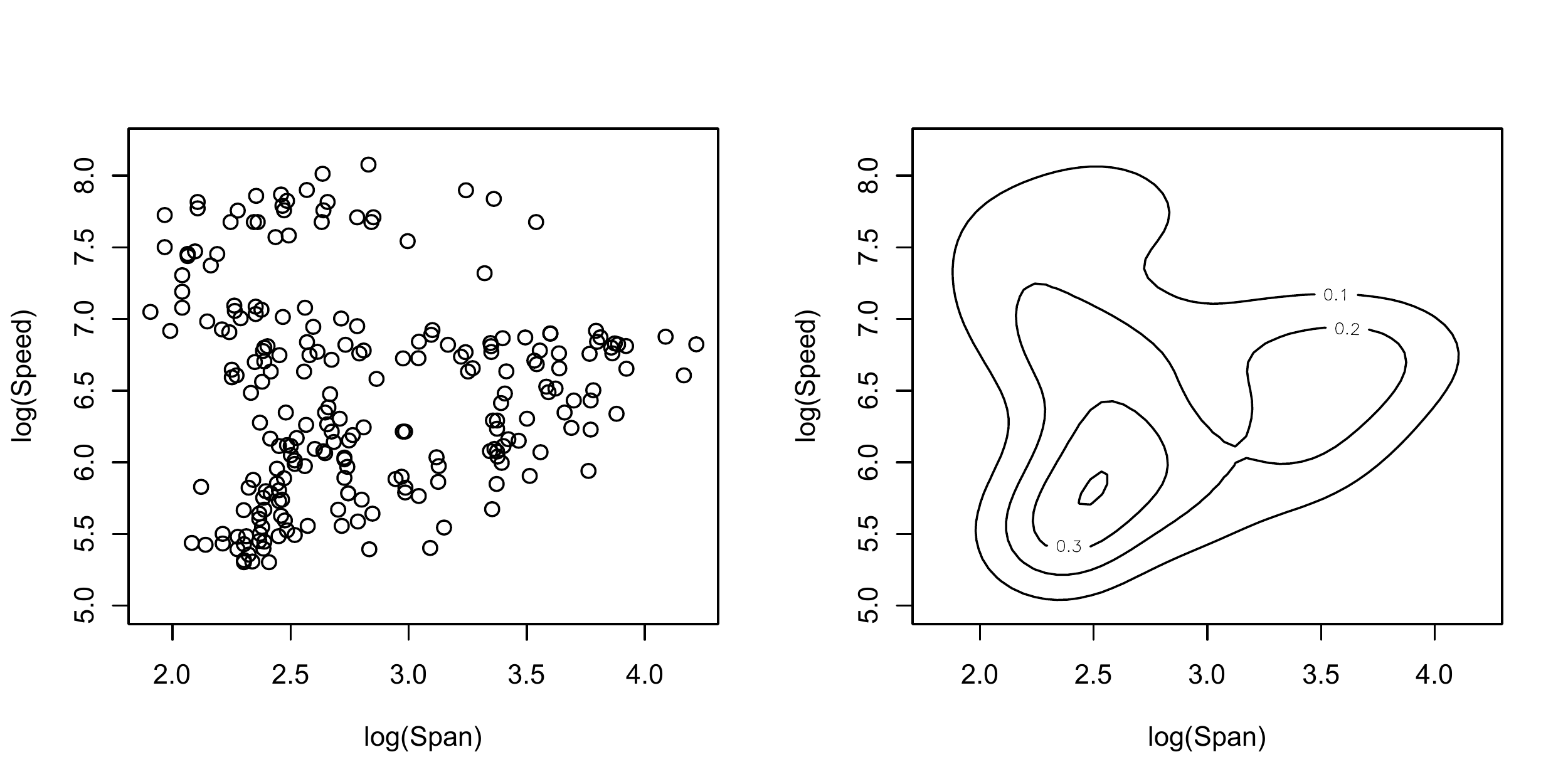}
                \caption{Showing the scatter plot and the density estimate contours for the aircraft speed and wing span} 
                \label{fig:5} 
                       
\end{figure}

We have explored the Saviotti aircraft data \citep*{saviotti1996} which was also analyzed by \citet*{szekely2009}. We consider the wing span (m) vs. speed (km/h) ($n=230$, \citet*{bowman1997}). \autoref{fig:5} shows the scatter plot of the two variables, alongside non-parametric density estimate contours (log scale). It is clear from the plot that there is a non-linear relationship ( Pearson's product moment correlation is a modest $0.0168$, $p$-value$=0.8001$), although the relationship is complicated and apparently not monotone.

All of the methods described here were significant at $\alpha=0.05$.  %[IS THAT WHAT YOU USED?]
The $p$-values for dCor, MIC, and HHG were  0.00013, 0.00004, and  $<10^{-5}$, respectively. For RankCover the test was also significant with a $p=0.0008$, and for the hybrid method $p= 0.0002$.
% 100000 simulations/permutations are used for each test. 

\subsubsection{Example 3: ENSO data}
\label{sec:enso}

The ENSO data ( also taken from the NIST Statistical Reference Datasets for non-linear regression)
consists of monthly average atmospheric pressure differences between Easter Island and Darwin, Australia \citep*{kahaner1989}, with 168 observations.
There are 168 observations.The data form a time series, and has different cyclical components which were modeled (\href{http://www.itl.nist.gov/div898/strd/nls/data/enso.shtml}{NIST StRD for non-linear regression}) by the proposed model

\begin{center}
$y=\beta_1+\beta_2 cos(\frac{2\pi x}{12})+\beta_3 sin(\frac{2\pi x}{12})
+\beta_5 cos(\frac{2\pi x}{\beta_4})+\beta_6 sin(\frac{2\pi x}{\beta_4})
+\beta_8 cos(\frac{2\pi x}{\beta_7})+\beta_9 sin(\frac{2\pi x}{\beta_7})
+\epsilon$,
\end{center}

where $\beta_1,\beta_2,...,\beta_9 \in \mathbb{R}$ and $\epsilon$ is random Gaussian noise.

\begin{figure}[!ht]
                \centering
                \includegraphics[width=10.5cm]{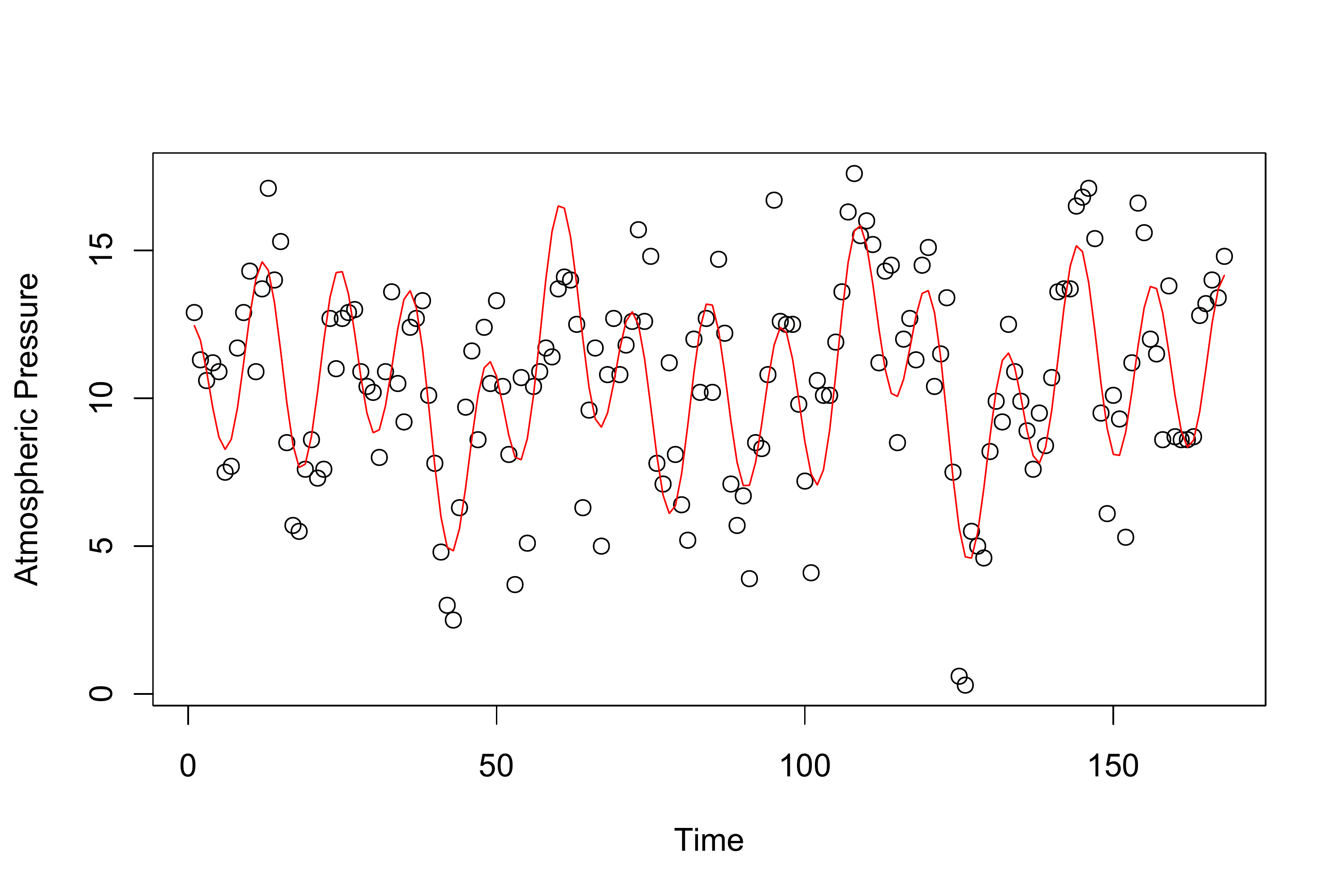}
                \caption{Showing the scatter plot and the fitted curve for the ENSO dataset} 
                \label{fig:7} 
                       
\end{figure}

\autoref{fig:7} shows the scatter plot of the data along with the fitted curve. The cyclical fluctuations are evident, but no
linear trend is observed. Thus, the Pearsonian correlation (0.0843) fails to capture the pattern. However a simple serial correlation with lag 1 (0.6102) reveals the association. With 100,000 simulations, the RankCover test is significant with $p$-value 0.00032. The hybrid test and MIC test are also significant with $p$-values 0.00064 and 0.00027 respectively. However dCor and HHG fail to detect significant association ($p$-values 0.13521 and 0.07617, respectively).

\subsubsection{Example 4: Yeast data}
\label{sec:yeast}

In this example, we analyze a yeast cell cycle gene expression dataset with 6223 genes \citet{spellman1998}.  The experiment was designed to identify genes with activity varying throughout the cell cycle \citep*{spellman1998}, and thus transcript levels would be expected to oscillate. This data has been analyzed by many researchers, including \citet*{reshef2011}, who used it to verifying the ability of MIC to detect oscillating patterns. We have run dCor, MIC, HHG, RankCover and the hybrid methods of test on the data and used the Benjamini-Hochberg method to control the false discovery rate. 

We have listed the genes identified by different methods after controlling the false discovery rate (FDR)  at the 5\% level and compared them with the list of genes identified by \citet*{spellman1998}. Of all the genes identified by \citet{spellman1998}, RankCover found 16\% to be significant, while dCor, MIC and HHG found only 6\%, 2\% and 8\% respectively. The hybrid method could identify 12\% of those genes.
% As mentioned in \citet{reshef2011}, such low percentage of genes is identified due to the small sample size (24) and for controlling FDR strictly. 
Instead controlling the FDR at 25\%, the figures for HHG, dCor, MIC, RankCover and the hybrid method become 39\%, 23\%, 18\%, 57\% and 47\% respectively. These figures differ slightly from those reported in \citet{reshef2011}, due to the difference in the procedure of handling the missing data (See Supplementary Article for details). 

\begin{figure}[!ht]
                \centering
                \includegraphics[width=\textwidth]{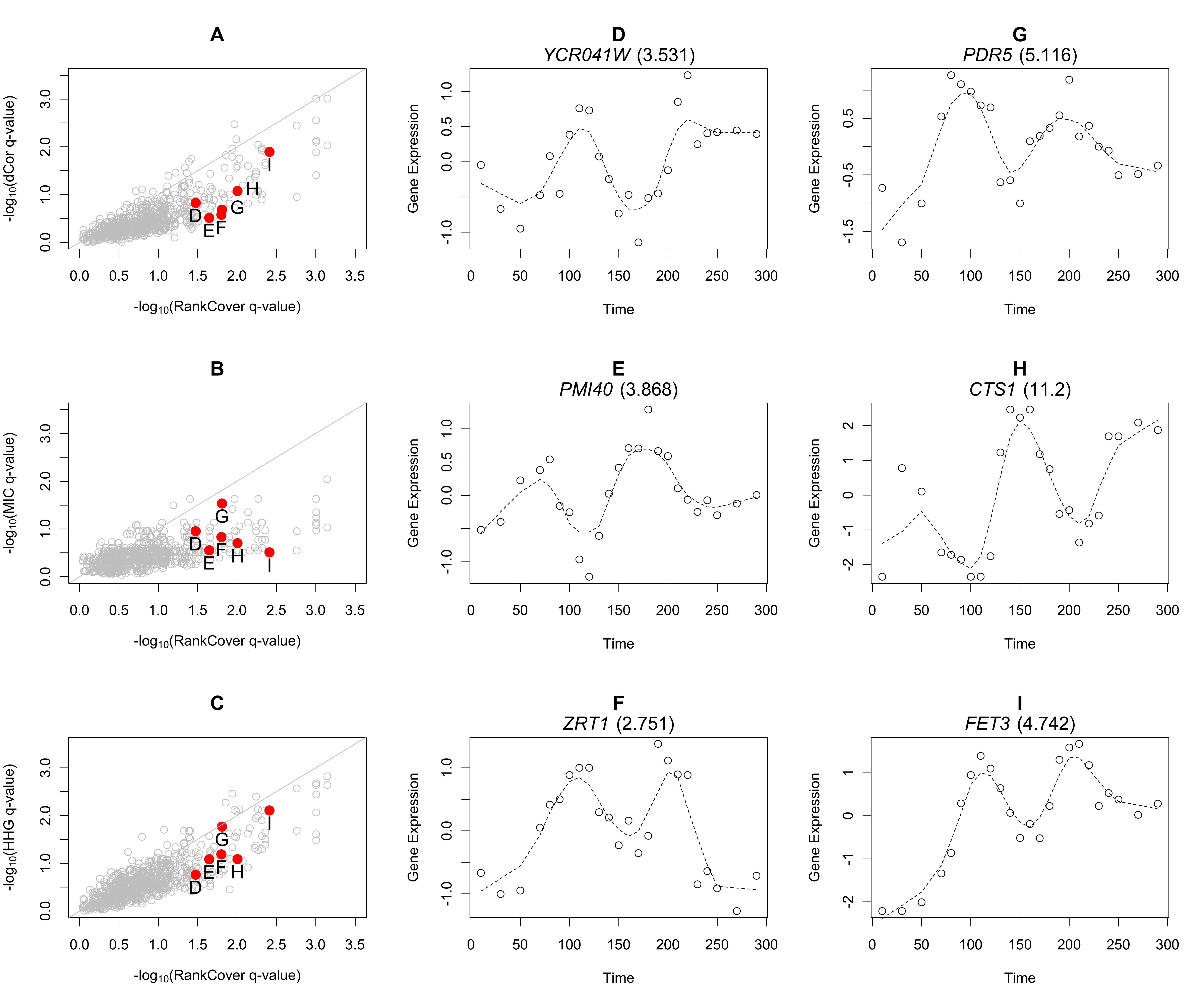}
                \caption{\textbf{A.} The plot comparing the FDR adjusted $q$-values of the test using RankCover and that using dCor for the genes in Spellman's list in a log scale. It is evident that most of the genes in Spellman's list have a smaller $q$-value when RankCover test is used. \textbf{B.} A similar plot comparing the $q$-values of RankCover and MIC. \textbf{C.} A similar plot comparing the $q$-values of RankCover and HHG. \textbf{D-I.} Examples of genes in the Spellman's list that were identified by RankCover, but not by at least one of dCor, MIC or HHG. The values in parentheses are the Spellman scores for the genes.} 
                \label{fig:8} 
                       
\end{figure}

For these data,  RankCover is clearly successful at identifying oscillating patterns expected for the experiment. This is also clear from \autoref{fig:8} (panel A, B and C) which compares the FDR adjusted $q$-values of our RankCover test with those of dCor, MIC and HHG on a logarithmic scale. Most of the genes in Spellman's list which are identified by dCor, MIC or HHG are also identified by RankCover, but RankCover identified more genes than the other methods. \autoref{fig:8} (panels D-I) shows some of the genes that are found significant by RankCover at 5\% level, but not found significant by at least one of the other three methods. {\it PDR5} was found significant by MIC, HHG and RankCover, but not by dCor. On the other hand MIC could not identify {\it FET3}, which was identified by dCor, HHG and RankCover. The other four genes shown in \autoref{fig:8} are found significant by RankCover but not by dCor, MIC or HHG. Note that all the six genes are found to be significant by the hybrid method.

\section{Summary}
\label{sec:summary}

Our RankCover testing procedure serves as a simple and powerful method to test for general association between a pair of variables. The method is applicable to the problem of testing general association irrespective of the marginal distributions of the (continuous) variables. Use of the rank scale also allows a  pre-computed null distribution for the statistic, avoiding the need for actual permutation. This, along with the introduction of the idea of using a single disk size makes the procedure computationally feasible. The testing procedure has been shown to be powerful in simulated datasets even with a small sample size. A variety of real datasets, ranging from studies of cell cycle effects in gene expression to studies involving circular interference transmittance show that the approach provides useful and interpretable results.

%RankCover doesn't have the nice theoretical properties like dCor, but its local nature of measuring association gives it high power against different forms of relationships which dCor fails to %identify.
Although dCor is theoretically motivated by consideration of characteristic functions, in practice it suffers for non-monotone relationships.
Our RankCover procedure is generally powerful and robust, and is more powerful than MIC, dCor and HHG for a number of scenarios.
RankCover may be especially useful to detect oscillating relationships, keeping in mind that such relationships need not be periodic and the amplitudes may vary.
A hybrid of RankCover and dCor is proposed, which is shown to be highly robust against many forms of associations.

%A procedure to test the association of two variables after removing the effect of a third variable is also introduced. With simulated data, we have illustrated that with an appropriate choice %of number of strata, the type-I error is controlled and reasonable power can be achieved at the same time. Though the choice of the number of strata depends on several factors and %therefore it is not possible to recommend a general choice, the result of our analysis provides some idea about it. This procedure is valid for any form of relationship between the three %variables and hence can be very useful in situations where the associations are not necessarily linear.

With the rapid rise of large datasets in today's scientific community, RankCover provides a useful tool to detect general association.  The approach is both sensitive and relatively powerful, even with small samples, against various and general forms of association.

%\begin{supplement}[id=suppA]
  %\sname{Supplementary Article}
  %\stitle{Title}
  %\slink[url]{www.google.com}
  %\sdatatype{.pdf}
  %\sdescription{Please read the file Supplementary.Article.pdf for details of our analysis.}
%\end{supplement}

\bibliography{rankcover}

\end{document}

\maketitle

%Details of our method%
%%%%%%%%%%%%%%%%%%%%%%%

\section{Details of the testing procedure}
\label{sec:details}

Let $(x_{k},y_{k})$ denote the ranks of the $k$th sample pair, $k=1,2,...,n$. We define

$d(i,j,x_k,y_k)$ = Distance between the point $(i,j)$ on the grid and $(x_k,y_k)$; 
\indent $d_{ij}=\min_{k} d(i,j,x_k,y_k)$.

\noindent The RankCover method measures the concentration of ranks using the test statistic

\begin{equation}
\label{eq:1}
T(\delta) = \frac{1}{n^2}\sum_{i=1-\lceil \delta \rceil}^{n+\lceil \delta \rceil} \sum_{j=1-\lceil \delta \rceil}^{n+\lceil \delta \rceil} {I(d_{ij} \leq \delta)},
\end{equation}

\noindent where $\delta$ is the disc radius (for Manhattan distance, $\delta$ is half the diagonal of each square). $\lceil t \rceil$ denotes the smallest integer greater than or equal to $t$.

\begin{figure}[!ht] 
                \centering
                \includegraphics[width=0.35\textwidth]{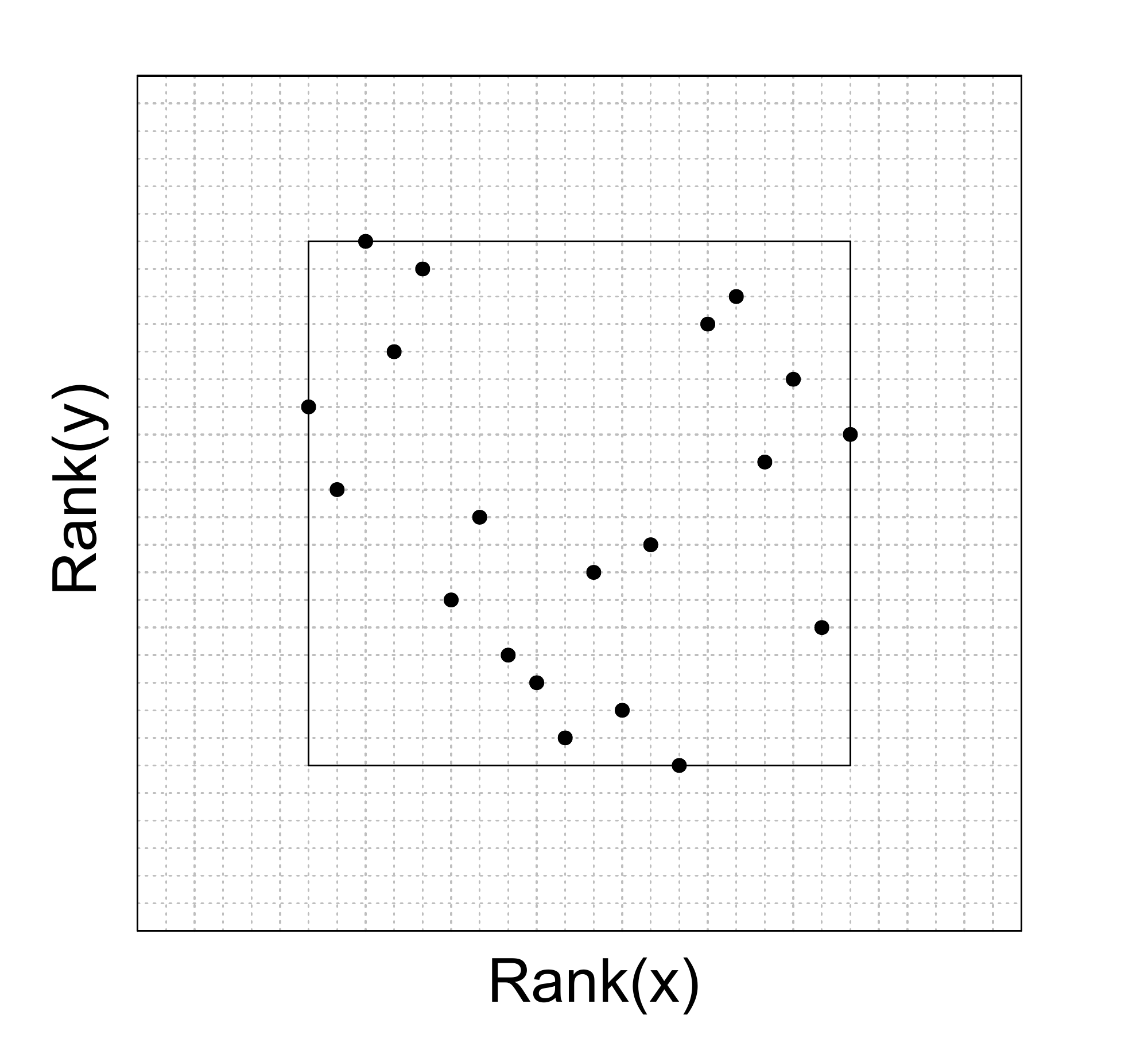} 
                \caption{Figure showing the grid used to calculate RankCover statistic: $n=20$, $ \lceil \delta_{opt} \rceil=6$, all the points on the grid are used to calculate the statistic rather than using just the points in the inner square}     
                \label{fig:s1} 
\end{figure}

An $(n+2\lceil \delta \rceil) \times (n+2\lceil \delta \rceil)$ grid is considered which is an outward extension of the $n \times n$ grid $\{1,2,...,n \} \times \{1,2,...,n \}$ (\autoref{fig:s1}).

%This is different from Diggle's F-function since we consider this outer region to take care of the edge effect. 

In order to do the test, one can pre-compute the threshold based on a large number of simulations. The use of ranks enables such pre-computation as the distribution of our test statistic under null doesn't depend on the distributions of $x$ and $y$. In \autoref{sec:tables}, we have presented a table of such pre-computed thresholds for some sample sizes.

%Justification for disc size%
%%%%%%%%%%%%%%%%%%%%%%%%%%%%%

\section{Choice of the disk size}
\label{sec:discsize}

The choice of the disc size $\delta$ is an important consideration. We have proposed the use of a single optimum choice of $\delta$ as opposed to the whole $\delta$ versus $F(\delta)$ curve used by Diggle. The argument for choosing $\delta_{opt}= \sqrt{n}$ for Euclidean distance and $\delta=\frac{\pi}{2}\sqrt{n}$ is somewhat heuristic, but based on empirical observations for several sample sizes.  

\begin{figure}[!ht] 
                \centering
                \includegraphics[width=0.6\textwidth]{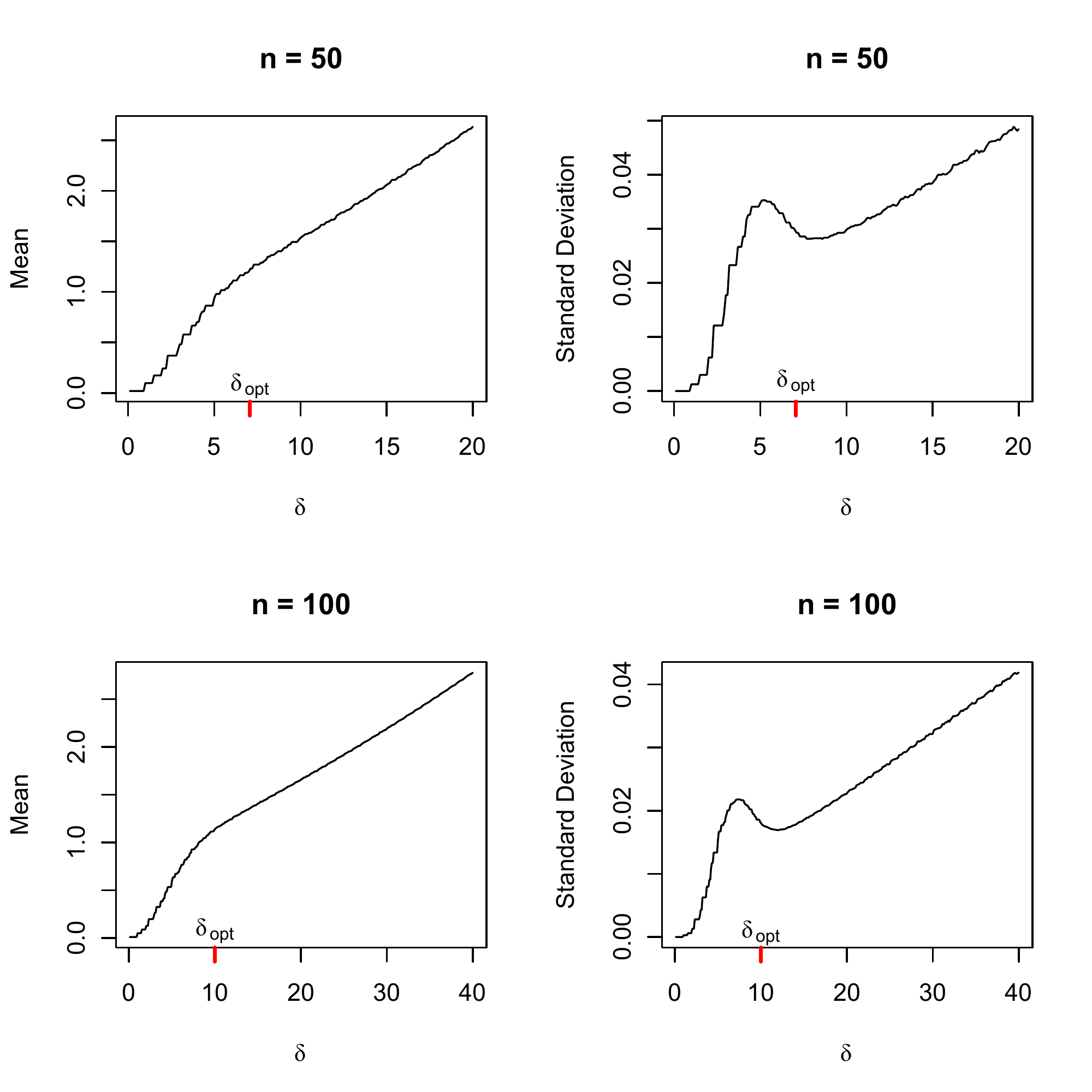} 
                \caption{Showing the mean and sd of $T(\delta)$ for sample sizes 50 and 100 (Euclidean distance is used)}     
                \label{fig:s2} 
\end{figure}

\begin{figure}[!ht] 
                \centering
                \includegraphics[width=0.6\textwidth]{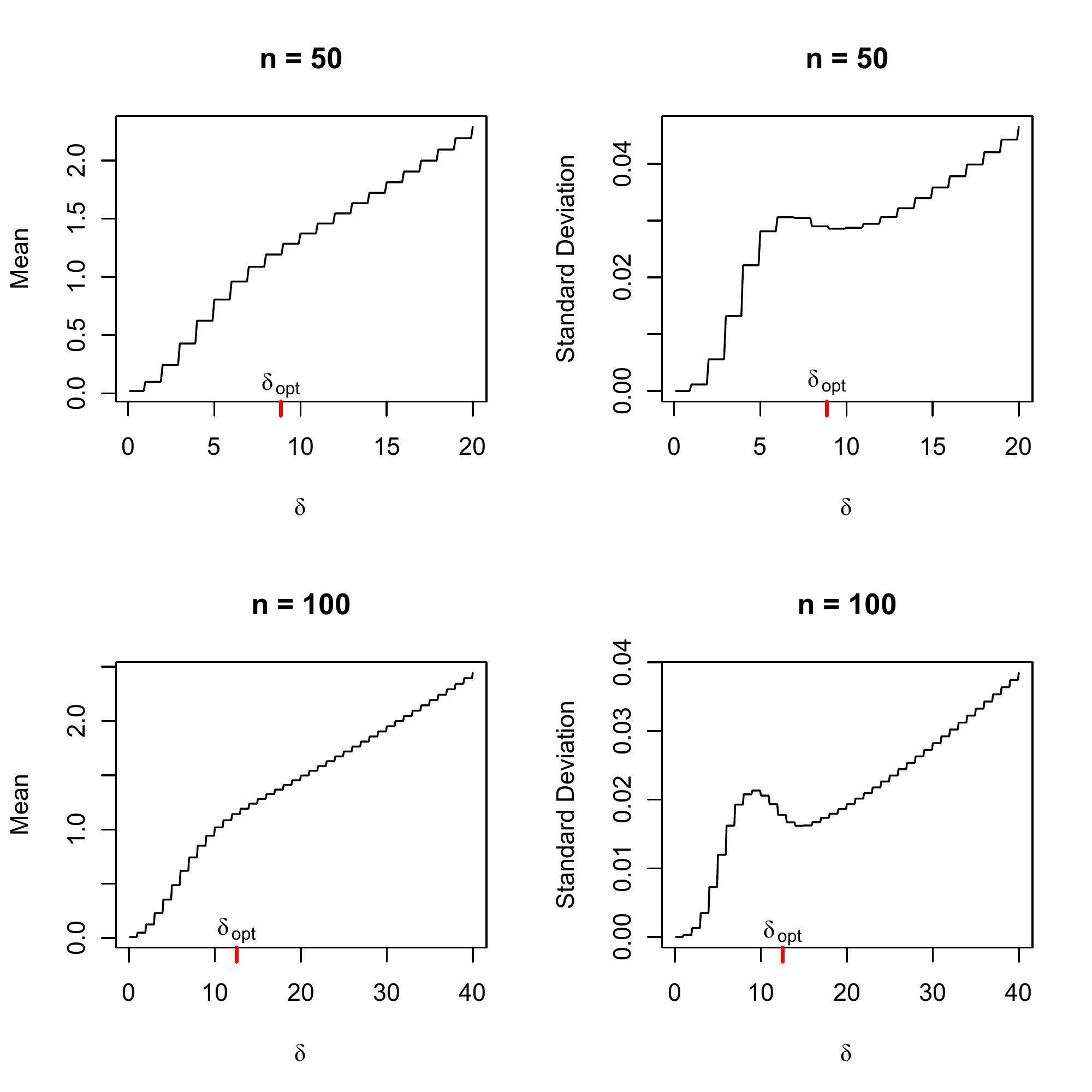} 
                \caption{Showing the mean and sd of $T(\delta)$ for sample sizes 50 and 100 (Manhattan distance is used)}     
                \label{fig:s3} 
\end{figure}

To understand the idea, we examine the expectation and standard deviation of $T(\delta)$ under null for varying $\delta$. These curves calculated based on 1000 simulations under null are shown in \autoref{fig:s2} for Euclidean distance and \autoref{fig:s3} for Manhattan distance. There is a clear change of curvature in the expectation in the vicinity of $\delta=\sqrt{n}$, and also we note that the standard deviation exhibits a local maximum and minimum in the vicinity. We reason that the local minimum of the standard deviation represents a good choice for $\delta$. We also note that the point where the expectation curve changes the curvature is approximately the same point as the local minimum of the standard deviation. However, there is no closed form expression for this point of local minimum. From simulations under different sample sizes, we have established that such local minima occur near $\delta= \sqrt{n} $ for Euclidian distance, and propose it as our choice of $\delta_{opt}$.  Also, it is clear from these simulations that if the distance metric is symmetric (eg Euclidian, Manhattan etc), the shape of these curves depend on $\delta$ only through the area of the disk, and so we use $\delta_{opt}=\sqrt{\frac{\pi}{2}n}$ for the Manhattan distance.

%Here we present similar plots like Figure 3 in the article for different sample sizes. \autoref{fig:s2} shows the plots for Euclidean distance, \autoref{fig:s3} shows the plots for Manhattan distance. 

%Is a single value of delta as good as the whole curve?%
%%%%%%%%%%%%%%%%%%%%%%%%%%%%%%%%%%%%%%%%%%%%%%%%%%%%%%%

\section{A single $\delta_{opt}$ vs. the entire curve}
\label{sec:singledelta}

\begin{figure}[!ht] 
                \centering
                \includegraphics[width=\textwidth]{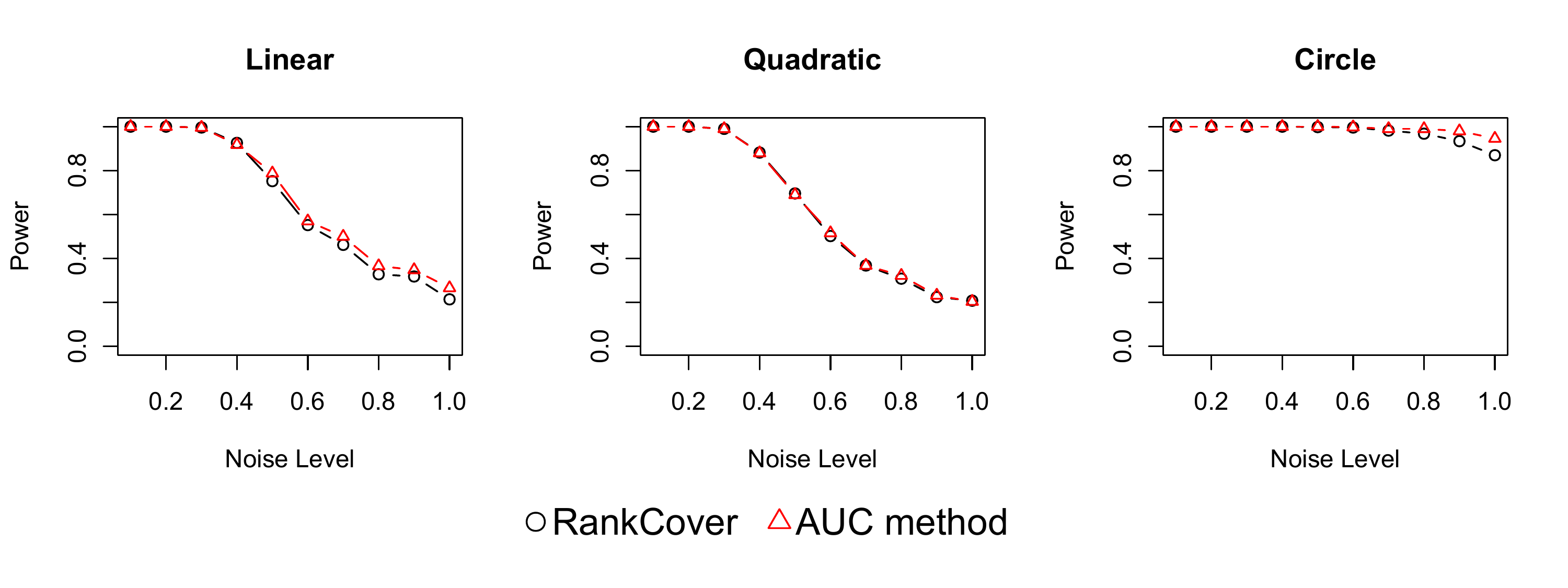} 
                \caption{Showing the power comparison of RankCover using $\delta_{opt}$ and the Area Under Curve method for three different types of relationship}     
                \label{fig:s4} 
\end{figure}

\autoref{fig:s4} shows an illustrative power comparison of our approach using a single optimum value of $\delta$ and the approach using the whole $\delta$ versus $F(\delta)$ curve. The second approach uses the area under curve as the test statistic. We have demonstrated the power comparisons for three different types of relationships: linear, quadratic and circular. It is clear from \autoref{fig:s4} that the use of a single $\delta$ doesn't reduce power substantially, but greatly reduces computation time.

%Details of simulation procedure%
%%%%%%%%%%%%%%%%%%%%%%%%%%%%%%%%%

\section{Details of the analysis of simulated data}
\label{sec:simulation}

This section explains the details of the analysis of simulated data in Section 3.1. We have used Manhattan distance throughout all the analyses due to the ease of tail area computation (\autoref{sec:tables}). RankCover procedure with Manhattan distance appears to give similar results to that with Euclidean distance (See \autoref{sec:distance}). 

The sample size is 50 (for other sample sizes see \autoref{sec:othern}) and we used 1000 simulations under the null for RankCover and MIC. For dCor and HHG, 1000 permutations are used. The power curves are obtained based on 500 simulations. The independent variable $x$ is simulated as $U(0,1)$. The dependent variable $y$ is calculated using the equation

\begin{equation}
y=f(x)+ \nu \times error,
\label{eq:2}
\end{equation}

where $\nu$ is the noise scale parameter and increases from 0.1 to 1 as in Figure 4.
The error distribution was chosen to be normal. However, as in \citet*{simon2014}, the variance of the error distribution was considered differently for different forms of relationship. \autoref{sec:otherdist} shows how the results are similar with other distributions also. The details of the forms of the function $f(.)$ and the error distributions are as below.

\begin{itemize}
\item Linear: $f(x)=x$ , error distribution is $N(0,1)$

\item Quadratic: $f(x)=4(x-1/2)^2$ , error distribution is $N(0,1)$

\item Cubic: $f(x)=128(x-1/3)^3-48(x-1/3)^2-12(x-1/3)$ , error distribution is $N(0,100)$

\item Sine: $f(x)=sin(4\pi x)$ , error distribution is $N(0,4)$

\item $X^{1/4}$: $f(x)=x^{1/4}$ , error distribution is $N(0,1)$

\item Circle: $f(x)=(2r-1)\sqrt{1-(2x-1)^2}$ , error distribution is $N(0,1/16)$, where $r$ is a Bernoulli($1/2$) variable

\item Two curves: $f(x)=2rx + (1-r)\sqrt{x}/2$ , error distribution is $N(0,1/4)$, where $r$ is a Bernoulli($1/2$) variable

\item X-function: $f(x)=rx + (1-r)(1-x)$ , error distribution is $N(0,1/25)$, where $r$ is a Bernoulli($1/2$) variable

\item Diamond: $f(x)= r_1 I(x<0.5) + r_2 I(x \geq 0.5)$ , error distribution is $N(0,1/100)$, where $r_1$ is a $U(0.5-x,0.5+x)$ variable and $r_2$ is a $U(x-0.5,1.5-x)$ variable

\end{itemize}

%Choice of distance metric%
%%%%%%%%%%%%%%%%%%%%%%%%%%%

\begin{figure}[!ht] 
                \centering
                \includegraphics[width=0.9\textwidth]{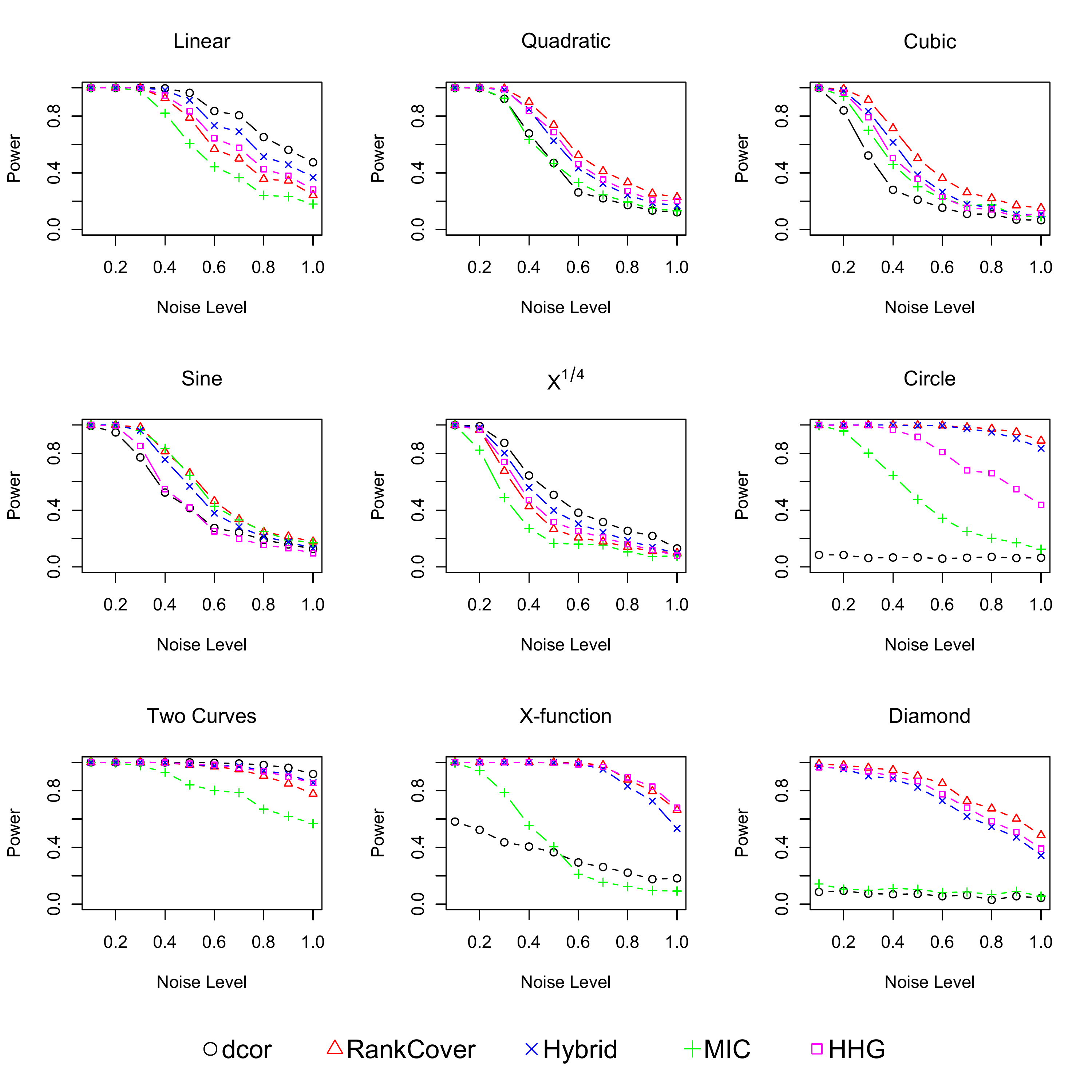} 
                \caption{Showing the power of different methods for Euclidean distance, $n=50$}     
                \label{fig:s5} 
\end{figure}

\section{Choice of distance metric}
\label{sec:distance}

We have explored two distance metrics: Euclidean distance and Manhattan distance. The performance of RankCover does not vary much based on the choice of the distance metric. \autoref{fig:s5} shows the power analysis on simulated data using Euclidean distance. The results are not much different from those obtained using Manhattan distance (Figure 4). However, we recommend Manhattan distance since it has the advantage of more easily approximating the tail area (\autoref{sec:tables}).

%Comparison with Spearman%
%%%%%%%%%%%%%%%%%%%%%%%%%%

\begin{figure}[!ht] 
                \centering
                \includegraphics[width=0.9\textwidth]{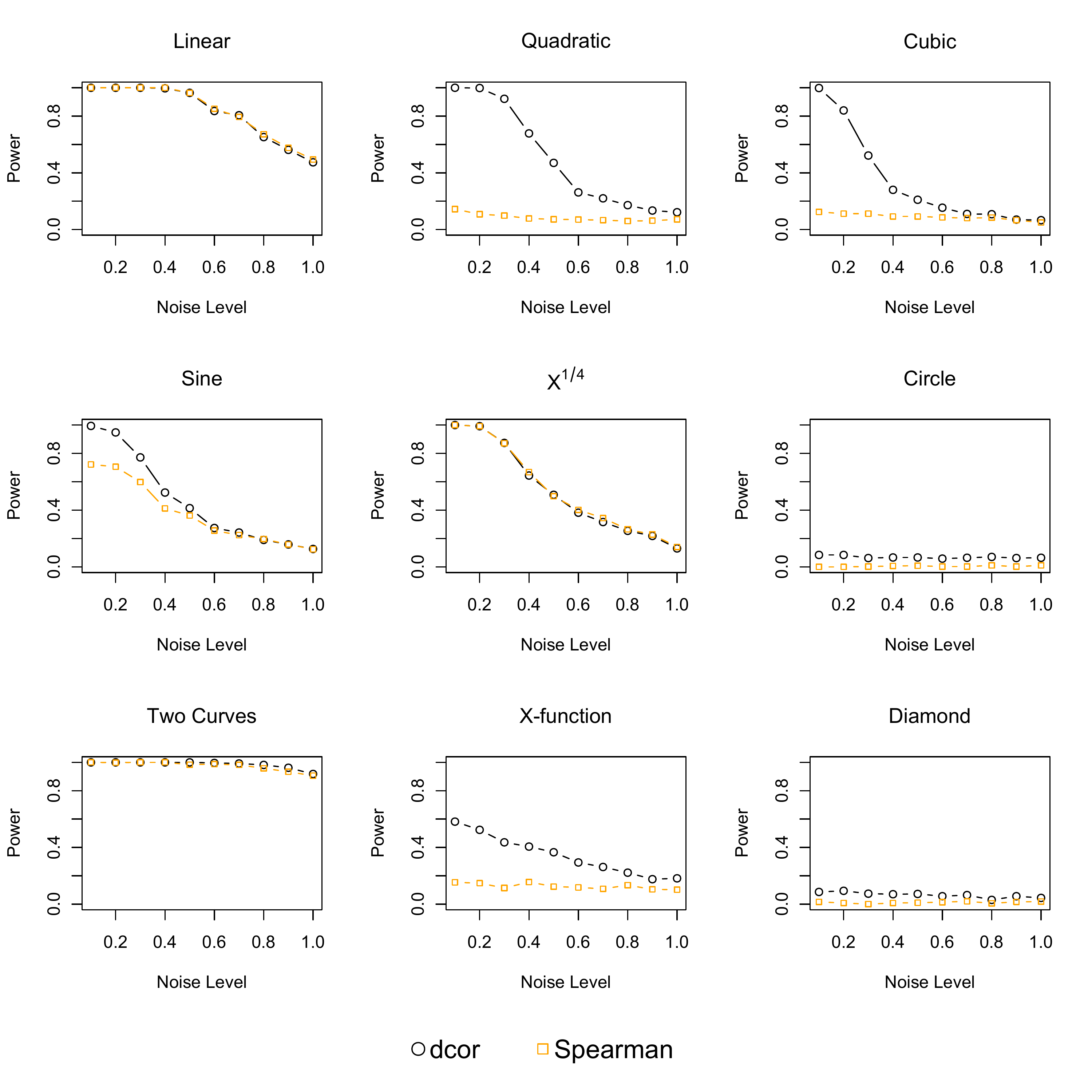} 
                \caption{Showing the power comparison of dCor and Spearman's rank correlation}     
                \label{fig:s5prime} 
\end{figure}

\section{Comparison of dCor with Spearman's rank correlation}
\label{sec:alsospearman}

The only cases where RankCover is dominated by some other method are all monotone relationships (linear, $X^{1/4}$, Two curves) and in all those cases dCor appears to be the best choice. However, we have shown (\autoref{fig:s5prime}) that even Spearman's rank correlation is equally powerful in those cases. Therefore, if we have prior knowledge that the relationship is monotone, then we do not gain anything by using the fancier methods anyway, and could use Spearman's rank correlation instead. We note that Spearman's rank correlation does not have much ``generality'' in the sense that it is not powerful against non-monotone alternatives. However, dCor has also been shown to have similar limitations.

%Other sample sizes%
%%%%%%%%%%%%%%%%%%%%

\section{Simulation results for some other sample sizes}
\label{sec:othern}

\begin{figure}[!ht] 
                \centering
                \includegraphics[width=0.9\textwidth]{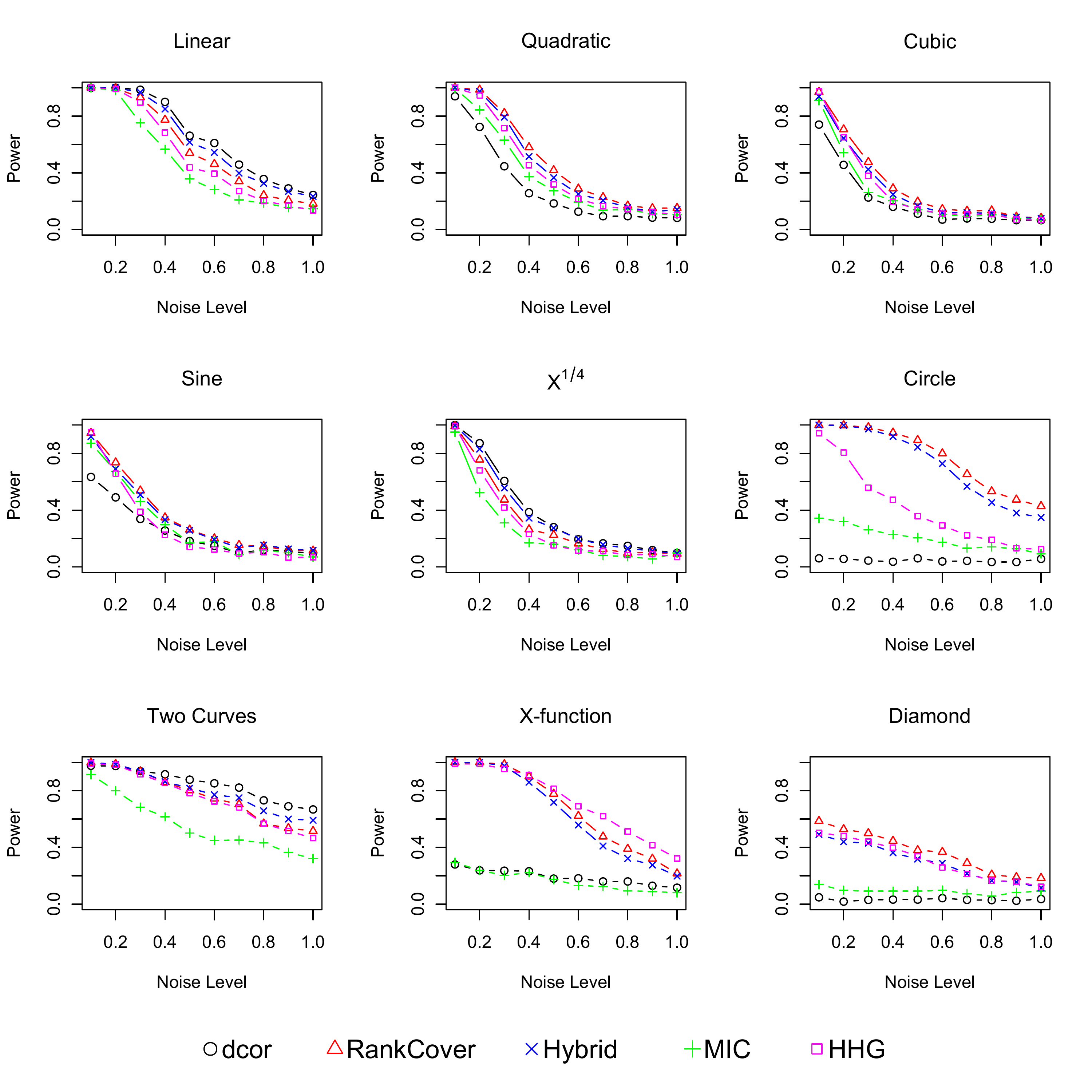} 
                \caption{Showing the power of different methods for $n=25$ (Manhattan distance)}     
                \label{fig:s6} 
\end{figure}

\begin{figure}[!ht] 
                \centering
                \includegraphics[width=0.9\textwidth]{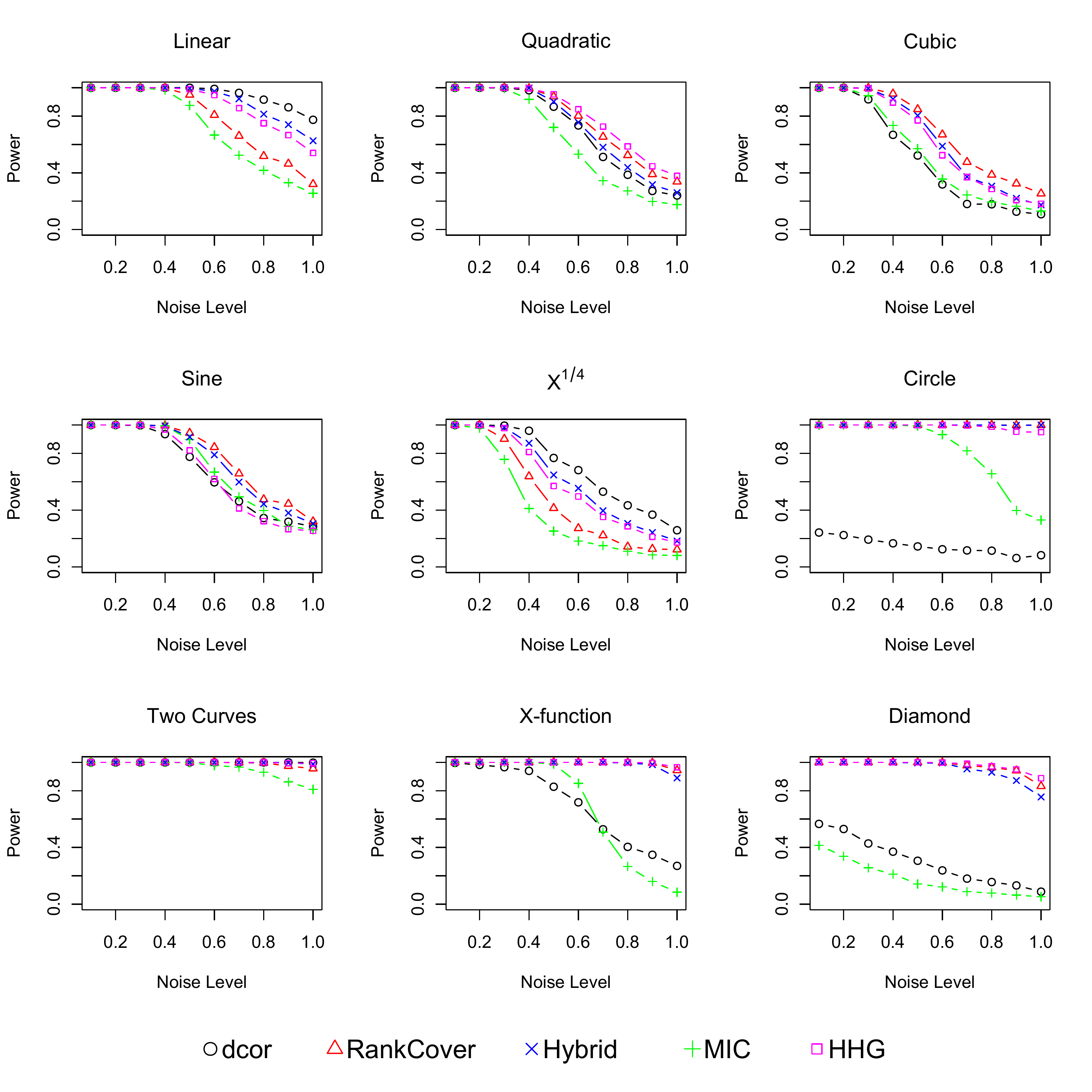} 
                \caption{Showing the power of different methods for $n=100$ (Manhattan distance)}     
                \label{fig:s7} 
\end{figure}

\autoref{fig:s6} and \autoref{fig:s7} show simulation results based on sample sizes 25 and 100 respectively. As an augment to Figure 4, the usefulness of RankCover is reflected in these figures.

%Other distributions of x and y%
%%%%%%%%%%%%%%%%%%%%%%%%%%%%%%%%

\section{Simulation results for different marginal distributions of $x$ and $y$}
\label{sec:otherdist}
We have carried out the simulation analysis for different marginal distributions of $x$ and different error distributions. Three distributions of different shapes are used for the marginal distribution of $X$: uniform, truncated normal (a normal distribution with mean $1/2$ and variance $1/12$ truncated between 0 and 1)and a U-shaped beta (beta($1/2,1/2$)). The choices for the error distributions are normal, U(0,1) and beta($1/2,1/2$) with appropriate shift of origin and scale so that the mean and variance of the error distributions are 0 and 1 respectively. 

\begin{figure}[!ht] 
                \centering
                \includegraphics[width=0.9\textwidth]{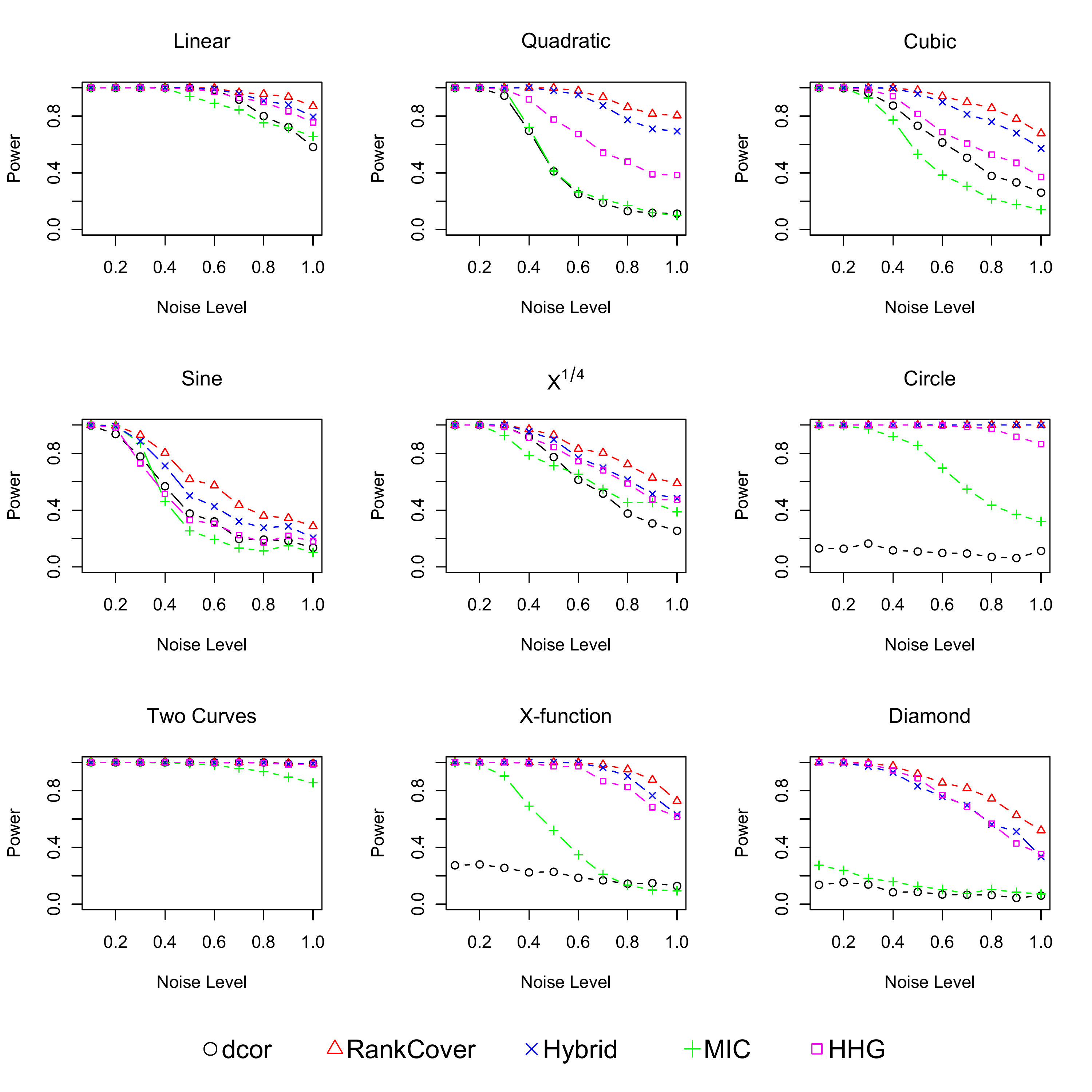} 
                \caption{Showing the power of different methods when marginal of $x$ is beta and error distribution is beta ($n=50$)}     
                \label{fig:s8} 
\end{figure}

\begin{figure}[!ht] 
                \centering
                \includegraphics[width=0.9\textwidth]{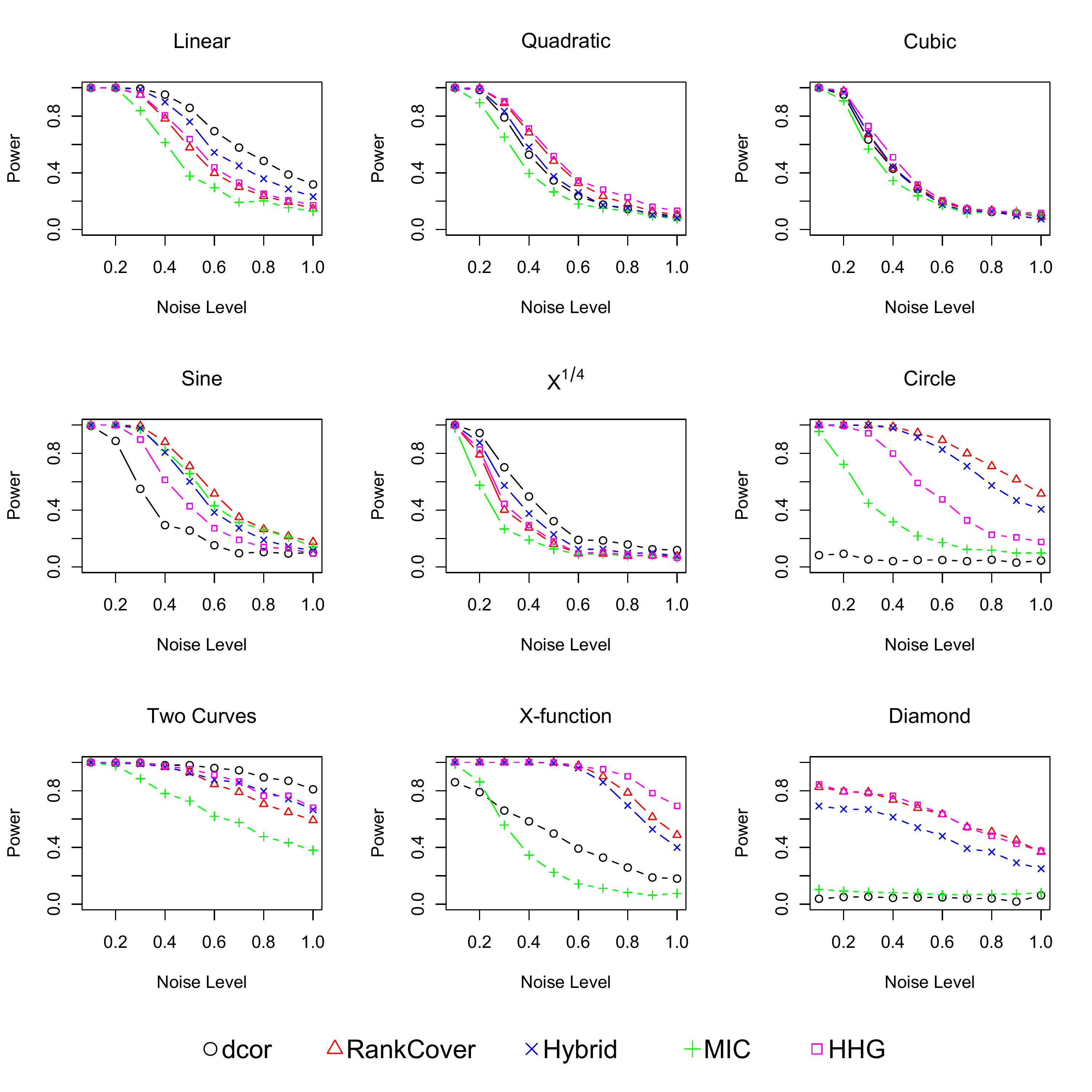} 
                \caption{Showing the power of different methods when marginal of $x$ is normal and error distribution is normal ($n=50$))}     
                \label{fig:s9} 
\end{figure}

The results of these nine cases show that RankCover has reasonable power in all these cases. It has very high power in some cases (\autoref{fig:s8}) and the power is not as high but still competitive in some other cases (\autoref{fig:s9}). \autoref{table:nineplot} shows a summary of all the cases. The mean power over all the noise levels are shown for each case. Since the power curves rarely cross each other, the mean power (which is approximately proportional to area under the power curve) appears to be a good indicator of performance.

%\begin{table}[!ht]

\tiny 
\renewcommand*{\arraystretch}{1.22}
%\centering
%\scalebox{0.7}{
\begin{longtable}{cccccccccc}
\caption{\textbf{Showing the mean power of the different methods for the nine cases. For each case, the first word is the distribution of $x$ and the second word is the error distribution. eg. Beta-Normal refers to the case where marginal of $x$ is beta and error distribution is normal}}\\

\hline
 & Linear & Quadratic & Cubic & Sine & $X^{1/4}$ & Circle & 2-Curves & X-function & Diamond \\ 
  \hline
  \endfirsthead

  \hline
 & Linear & Quadratic & Cubic & Sine & $X^{1/4}$ & Circle & 2-Curves & X-function & Diamond \\ 
  \hline
  \endhead
  
  \multicolumn{10}{r@{}}{continued to next page\ldots}\\
	\endfoot    
	\hline
	\endlastfoot
  
  \multicolumn{10}{c}{Beta-Beta}\\
  \hline
  dCor & 0.90 & 0.48 & 0.67 & 0.47 & 0.67 & 0.11 & 1.00 & 0.20 & 0.09 \\ 
	RankCover & 0.97 & 0.94 & 0.91 & 0.63 & 0.85 & 1.00 & 1.00 & 0.95 & 0.84 \\ 
	Hybrid & 0.95 & 0.90 & 0.87 & 0.56 & 0.79 & 1.00 & 1.00 & 0.93 & 0.76 \\ 
  MIC & 0.88 & 0.50 & 0.55 & 0.43 & 0.69 & 0.71 & 0.96 & 0.50 & 0.14 \\ 
  HHG & 0.94 & 0.72 & 0.74 & 0.47 & 0.77 & 0.97 & 1.00 & 0.89 & 0.76 \\

  \hline
  \multicolumn{10}{c}{Beta-Normal}\\
  \hline
  dCor & 0.90 & 0.52 & 0.69 & 0.51 & 0.69 & 0.10 & 1.00 & 0.19 & 0.09 \\ 
  RankCover & 0.75 & 0.72 & 0.75 & 0.48 & 0.54 & 1.00 & 0.97 & 0.94 & 0.81 \\ 
  Hybrid & 0.86 & 0.67 & 0.74 & 0.49 & 0.62 & 1.00 & 0.99 & 0.91 & 0.74 \\
  MIC & 0.70 & 0.61 & 0.61 & 0.51 & 0.46 & 0.73 & 0.93 & 0.51 & 0.15 \\ 
  HHG & 0.81 & 0.65 & 0.70 & 0.46 & 0.59 & 0.98 & 0.99 & 0.89 & 0.77 \\

  \hline
  \multicolumn{10}{c}{Beta-Uniform}\\
  \hline
  dCor & 0.89 & 0.49 & 0.67 & 0.46 & 0.66 & 0.11 & 1.00 & 0.20 & 0.09 \\ 
  RankCover & 0.85 & 0.80 & 0.81 & 0.50 & 0.63 & 1.00 & 0.99 & 0.94 & 0.83 \\ 
  Hybrid & 0.86 & 0.74 & 0.77 & 0.47 & 0.62 & 1.00 & 0.99 & 0.91 & 0.75 \\
  MIC & 0.74 & 0.51 & 0.56 & 0.44 & 0.47 & 0.71 & 0.93 & 0.50 & 0.15 \\ 
  HHG & 0.82 & 0.62 & 0.69 & 0.41 & 0.60 & 0.97 & 0.99 & 0.88 & 0.76 \\

  \hline
  \multicolumn{10}{c}{Normal-Beta}\\
  \hline
  dCor & 0.71 & 0.42 & 0.38 & 0.34 & 0.40 & 0.05 & 0.94 & 0.46 & 0.05 \\ 
  RankCover & 0.86 & 0.76 & 0.68 & 0.81 & 0.58 & 0.87 & 0.90 & 0.89 & 0.63 \\ 
  Hybrid & 0.81 & 0.68 & 0.59 & 0.72 & 0.51 & 0.82 & 0.90 & 0.84 & 0.49 \\ 
  MIC & 0.69 & 0.32 & 0.44 & 0.47 & 0.42 & 0.33 & 0.73 & 0.33 & 0.08 \\ 
  HHG & 0.77 & 0.64 & 0.55 & 0.50 & 0.49 & 0.57 & 0.90 & 0.92 & 0.64 \\

  \hline
  \multicolumn{10}{c}{Normal-Normal}\\
  \hline
  dCor & 0.73 & 0.44 & 0.40 & 0.35 & 0.42 & 0.05 & 0.94 & 0.47 & 0.04 \\ 
  RankCover & 0.56 & 0.50 & 0.41 & 0.61 & 0.30 & 0.85 & 0.85 & 0.88 & 0.63 \\ 
  Hybrid & 0.65 & 0.45 & 0.40 & 0.55 & 0.36 & 0.79 & 0.88 & 0.84 & 0.50 \\ 
  MIC & 0.48 & 0.38 & 0.37 & 0.58 & 0.25 & 0.33 & 0.69 & 0.35 & 0.08 \\ 
  HHG & 0.58 & 0.53 & 0.43 & 0.48 & 0.32 & 0.57 & 0.89 & 0.93 & 0.63 \\

  \hline
  \multicolumn{10}{c}{Normal-Uniform}\\
  \hline
  dCor & 0.70 & 0.41 & 0.37 & 0.34 & 0.40 & 0.06 & 0.93 & 0.47 & 0.05 \\ 
  RankCover & 0.65 & 0.57 & 0.49 & 0.69 & 0.36 & 0.84 & 0.85 & 0.87 & 0.62 \\ 
  Hybrid & 0.65 & 0.50 & 0.44 & 0.60 & 0.36 & 0.78 & 0.87 & 0.83 & 0.50 \\ 
  MIC & 0.50 & 0.33 & 0.35 & 0.50 & 0.26 & 0.33 & 0.67 & 0.33 & 0.08 \\ 
  HHG & 0.59 & 0.53 & 0.42 & 0.45 & 0.33 & 0.55 & 0.88 & 0.91 & 0.63 \\ 
  
  \hline
  \multicolumn{10}{c}{Uniform-Beta}\\
  \hline
  dCor & 0.82 & 0.47 & 0.32 & 0.44 & 0.51 & 0.06 & 0.98 & 0.32 & 0.07 \\ 
  RankCover & 0.93 & 0.88 & 0.74 & 0.76 & 0.71 & 0.98 & 0.98 & 0.94 & 0.77 \\ 
  Hybrid & 0.90 & 0.81 & 0.65 & 0.67 & 0.64 & 0.97 & 0.97 & 0.91 & 0.68 \\ 
  MIC & 0.79 & 0.42 & 0.38 & 0.46 & 0.55 & 0.50 & 0.87 & 0.42 & 0.10 \\ 
  HHG & 0.87 & 0.68 & 0.52 & 0.48 & 0.61 & 0.78 & 0.97 & 0.93 & 0.73 \\

  \hline
  \multicolumn{10}{c}{Uniform-Normal}\\
  \hline
  dCor & 0.81 & 0.46 & 0.27 & 0.43 & 0.51 & 0.12 & 0.97 & 0.26 & 0.04 \\
  RankCover & 0.66 & 0.62 & 0.51 & 0.59 & 0.39 & 0.98 & 0.94 & 0.93 & 0.78 \\ 
  Hybrid & 0.76 & 0.57 & 0.45 & 0.54 & 0.46 & 0.96 & 0.96 & 0.90 & 0.68 \\ 
  MIC & 0.59 & 0.51 & 0.42 & 0.58 & 0.33 & 0.50 & 0.82 & 0.44 & 0.09 \\ 
  HHG & 0.71 & 0.60 & 0.44 & 0.46 & 0.43 & 0.80 & 0.96 & 0.94 & 0.76 \\

  \hline
  \multicolumn{10}{c}{Uniform-Uniform}\\
  \hline
  dCor & 0.80 & 0.46 & 0.32 & 0.44 & 0.50 & 0.07 & 0.98 & 0.33 & 0.06 \\ 
  RankCover & 0.75 & 0.70 & 0.57 & 0.64 & 0.47 & 0.97 & 0.95 & 0.92 & 0.78 \\ 
  Hybrid & 0.76 & 0.63 & 0.50 & 0.58 & 0.46 & 0.95 & 0.96 & 0.89 & 0.68 \\ 
  MIC & 0.61 & 0.43 & 0.36 & 0.48 & 0.35 & 0.49 & 0.82 & 0.42 & 0.10 \\ 
  HHG & 0.71 & 0.59 & 0.43 & 0.44 & 0.43 & 0.76 & 0.96 & 0.92 & 0.74 \\

   \hline
\label{table:nineplot}

\end{longtable}
%}
\normalsize
%\end{table}

%Data analysis details%
%%%%%%%%%%%%%%%%%%%%%%%

\section{Details of real data analyses}
\label{sec:dataanalysis}

\subsection{Example 1: Eckerle4 data}
\label{sec:eckerle4}
100,000 simulations were used for RankCover and MIC. 100,000 permutations were used for dCor and HHG. The estimates of $\beta_1, \beta_2, \beta_3$ obtained from NIST website are used for plotting the fitted curve in Figure 5. 
Source of data: \href{http://www.itl.nist.gov/div898/strd/nls/data/eckerle4.shtml}{NIST StRD for non-linear regression}.

\subsection{Example 2: Aircraft data}
\label{sec:aircraft}

100,000 simulations were used for RankCover and MIC. 100,000 permutations were used for dCor and HHG. 
Source of data: \texttt{sm} Package in R \citep*{bowman2013r}.

\subsection{Example 3: ENSO data}
\label{sec:enso}

100,000 simulations were used for RankCover and MIC. 100,000 permutations were used for dCor and HHG. The estimates of $\beta_1, \beta_2,..., \beta_9$ obtained from NIST website are used for plotting the fitted curve in Figure 7. 
Source of data: \href{http://www.itl.nist.gov/div898/strd/nls/data/enso.shtml}{NIST StRD for non-linear regression}.

\subsection{Example 4: Yeast data}
\label{sec:yeast}

100,000 simulations were used for RankCover and MIC. 100,000 permutations were used for dCor and HHG. 
The data was pre-processed before analysis as follows. The data contained several missing observations. Since the sample size is small (24), we removed all the genes that had more than 3 missing observations. All other missing observations were imputed using KNN imputation \citep*{troyanskaya2001}. Then quantile normalization was used to normalize the data. Unlike \citet*{reshef2011}, we didn't remove any of the time points and didn't use any interpolation to find expression values for intermediate timepoints.  
Source of data: \href{http://www.molbiolcell.org/content/9/12/3273/suppl/DC1}{Comprehensive Identification of Cell Cycle regulated Genes of the Yeast Saccharomyces cerevisiae by Microarray Hybridization}.

%Stratified analysis details%
%%%%%%%%%%%%%%%%%%%%%%%%%%%%%

%\section{Details of stratified analysis}
%\label{sec:stratanalysis}

%To get the two variables $x$ and $y$ having a marginal relationship and also both of them having a dependence on a third variable $z$, we used the following simulation scheme. We simulated variables $x_1$ and $y_1$ just like \autoref{sec:simulation} with $x_1$ following $U(0,1)$ and $y_1$ determined by \autoref{eq:2}. $\nu$ is chosen to be $0.5$. Then, $z$ was also simulated from $U(0,1)$ and the two variables $x$ and $y$ were defined as

%\begin{center}
%$x=x_1+3z$\\
%$y=y_1+2z$
%\end{center}

%The power in the ideal situation is obatained by applying RankCover on $x_1$ and $y_1$. The inflation of the Type-I error is explored by testing on $x$ and $y$ when $x_1$ and $y_1$ are simulated from independent standard normal distributions.

%Tables of pre-computed thresholds%
%%%%%%%%%%%%%%%%%%%%%%%%%%%%%%%%%%%

\section{Tables of pre-computed thresholds}
\label{sec:tables}

\begin{figure}[!ht] 
                \centering
                \includegraphics[width=0.85\textwidth]{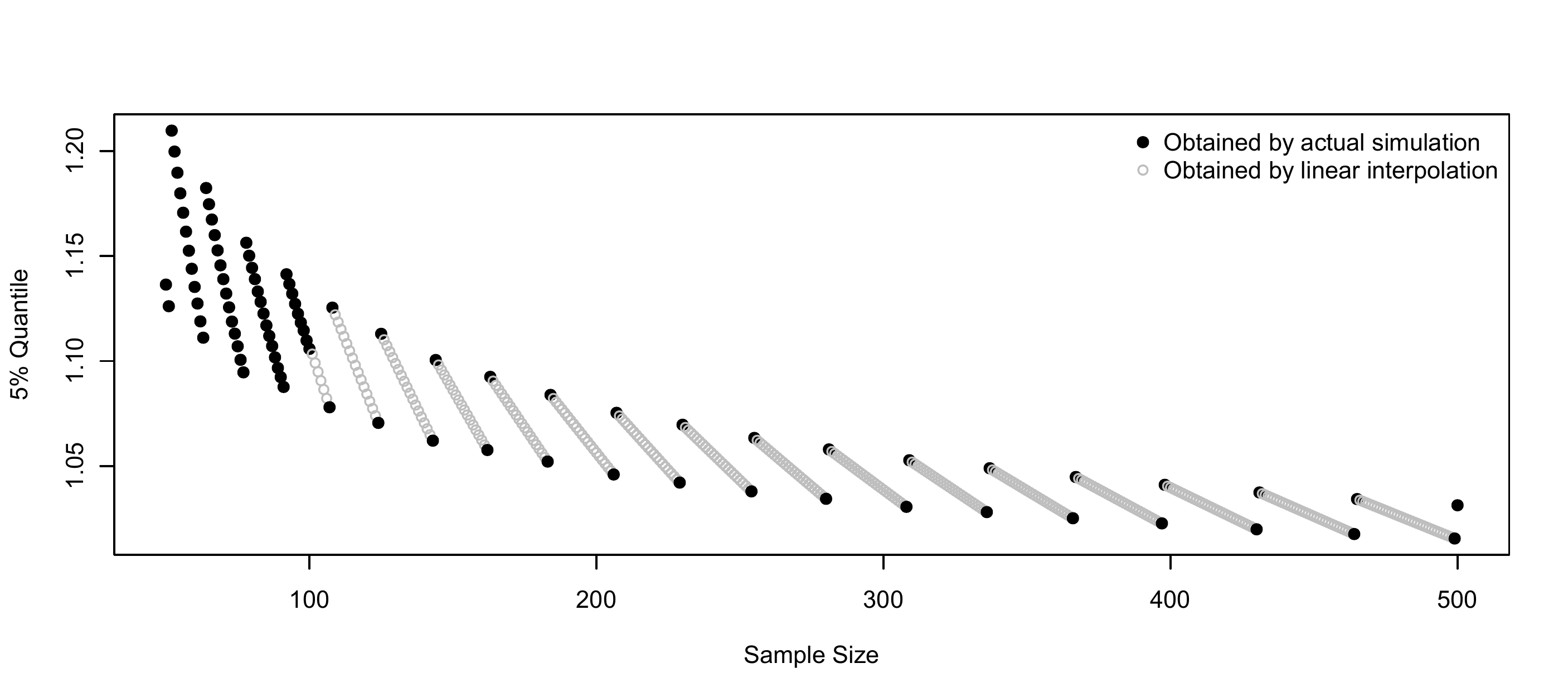} 
                \caption{Showing the 5th percentiles of the Rankcover statistic for Manhattan distance for $n=50,...,500$. Similar pattern is observed for other percentiles also.}     
                \label{fig:sawtooth} 
\end{figure}

The use of ranks in our procedure enables us to build tables of pre-computed thresholds for the
test. Such pre-computed thresholds for RankCover method with Manhattan distance are given in \autoref{table:s2} and those for the hybrid method are given in \autoref{table:s3}. 100000 simulations were used to  calculate the thresholds in each case.  For the Manhattan metric, the rejection thresholds follow a sawtooth pattern (\autoref{fig:sawtooth}), with jump points occurring at the values of $n$ where $[\delta]$ changes.  Simulations were performed for $n=20,...,100$.  For large values of $n$, to reduce computation, tables were generated by (1) performing direct simulation for the values of $n$ at, and just prior to, the jump points, followed by (2) linear interpolation for remaining values of $n$.

\clearpage

\tiny 
\renewcommand*{\arraystretch}{1.22}
\begin{longtable}{ccccccc}
\caption{\textbf{Showing the $p$-th quantiles of the RankCover statistic}} \\

  \hline
Sample Sizes & p=0.1 & p=0.05 & p=0.025 & p=0.01 & p=0.001 & p=0.0001 \\ 
  \hline
  \endhead
  \hline
\multicolumn{7}{r@{}}{continued to next page\ldots}\\
\endfoot    
\hline
\endlastfoot

			20 & 1.31000 & 1.28500 & 1.26000 & 1.23250 & 1.16500 & 1.10500 \\ 
      21 & 1.27211 & 1.24717 & 1.22449 & 1.19501 & 1.12925 & 1.07936 \\ 
      22 & 1.23554 & 1.21281 & 1.19215 & 1.16529 & 1.10331 & 1.06405 \\ 
      23 & 1.39698 & 1.37240 & 1.35161 & 1.32325 & 1.26087 & 1.20227 \\ 
      24 & 1.36111 & 1.33854 & 1.31771 & 1.29167 & 1.23090 & 1.16493 \\ 
      25 & 1.32960 & 1.30720 & 1.28640 & 1.26240 & 1.20640 & 1.14720 \\ 
      26 & 1.30030 & 1.27959 & 1.25888 & 1.23373 & 1.17899 & 1.12574 \\ 
      27 & 1.27298 & 1.25240 & 1.23320 & 1.20850 & 1.15501 & 1.10151 \\ 
      28 & 1.24745 & 1.22832 & 1.20918 & 1.18622 & 1.13520 & 1.08291 \\ 
      29 & 1.22473 & 1.20452 & 1.18668 & 1.16290 & 1.11415 & 1.07134 \\ 
      30 & 1.20222 & 1.18333 & 1.16444 & 1.14222 & 1.09222 & 1.04889 \\ 
      31 & 1.18106 & 1.16233 & 1.14464 & 1.12279 & 1.07700 & 1.03018 \\ 
      32 & 1.30469 & 1.28613 & 1.26855 & 1.24609 & 1.19922 & 1.15625 \\ 
      33 & 1.28375 & 1.26538 & 1.24885 & 1.22865 & 1.18182 & 1.14509 \\ 
      34 & 1.26384 & 1.24567 & 1.22924 & 1.20934 & 1.16263 & 1.11938 \\ 
      35 & 1.24490 & 1.22776 & 1.21061 & 1.19102 & 1.14286 & 1.10286 \\ 
      36 & 1.22685 & 1.20988 & 1.19367 & 1.17361 & 1.13040 & 1.09259 \\ 
      37 & 1.21110 & 1.19430 & 1.17823 & 1.15997 & 1.11395 & 1.08400 \\ 
      38 & 1.19453 & 1.17798 & 1.16274 & 1.14474 & 1.10319 & 1.06856 \\ 
      39 & 1.17883 & 1.16239 & 1.14727 & 1.12821 & 1.08613 & 1.05523 \\ 
      40 & 1.16375 & 1.14813 & 1.13375 & 1.11563 & 1.07437 & 1.04375 \\ 
      41 & 1.26413 & 1.24866 & 1.23379 & 1.21594 & 1.17668 & 1.14456 \\ 
      42 & 1.24943 & 1.23413 & 1.21995 & 1.20181 & 1.16213 & 1.12132 \\ 
      43 & 1.23580 & 1.22012 & 1.20606 & 1.18875 & 1.15035 & 1.10871 \\ 
      44 & 1.22159 & 1.20713 & 1.19318 & 1.17717 & 1.13998 & 1.10795 \\ 
      45 & 1.20889 & 1.19407 & 1.18074 & 1.16395 & 1.12444 & 1.09284 \\ 
      46 & 1.19660 & 1.18195 & 1.16824 & 1.15217 & 1.11531 & 1.08932 \\ 
      47 & 1.18470 & 1.17021 & 1.15708 & 1.14079 & 1.10457 & 1.07062 \\ 
      48 & 1.17231 & 1.15842 & 1.14497 & 1.12934 & 1.09115 & 1.06510 \\ 
      49 & 1.16160 & 1.14744 & 1.13369 & 1.11828 & 1.08330 & 1.05373 \\ 
      50 & 1.15080 & 1.13640 & 1.12400 & 1.10760 & 1.07360 & 1.04520 \\ 
      51 & 1.13995 & 1.12611 & 1.11342 & 1.09765 & 1.06267 & 1.03114 \\ 
      52 & 1.22337 & 1.20969 & 1.19749 & 1.18158 & 1.14756 & 1.11501 \\ 
      53 & 1.21289 & 1.19972 & 1.18726 & 1.17230 & 1.13884 & 1.11463 \\ 
      54 & 1.20302 & 1.18964 & 1.17764 & 1.16324 & 1.12929 & 1.10391 \\ 
      55 & 1.19273 & 1.17983 & 1.16793 & 1.15372 & 1.12066 & 1.09421 \\ 
      56 & 1.18367 & 1.17060 & 1.15912 & 1.14445 & 1.11129 & 1.07175 \\ 
      57 & 1.17421 & 1.16159 & 1.15020 & 1.13573 & 1.10249 & 1.07572 \\ 
      58 & 1.16498 & 1.15250 & 1.14090 & 1.12634 & 1.09304 & 1.06421 \\ 
      59 & 1.15628 & 1.14392 & 1.13243 & 1.11807 & 1.08475 & 1.05918 \\ 
      60 & 1.14778 & 1.13528 & 1.12361 & 1.11000 & 1.07805 & 1.05333 \\ 
      61 & 1.13948 & 1.12739 & 1.11610 & 1.10266 & 1.06987 & 1.04139 \\ 
      62 & 1.13137 & 1.11889 & 1.10744 & 1.09417 & 1.06322 & 1.03668 \\ 
      63 & 1.12371 & 1.11111 & 1.10028 & 1.08743 & 1.05694 & 1.03452 \\ 
      64 & 1.19434 & 1.18237 & 1.17188 & 1.15845 & 1.12524 & 1.10181 \\ 
      65 & 1.18627 & 1.17467 & 1.16426 & 1.15101 & 1.12260 & 1.09870 \\ 
      66 & 1.17906 & 1.16736 & 1.15657 & 1.14371 & 1.11524 & 1.08655 \\ 
      67 & 1.17153 & 1.15995 & 1.14970 & 1.13678 & 1.11049 & 1.08599 \\ 
      68 & 1.16436 & 1.15268 & 1.14208 & 1.12954 & 1.10208 & 1.07656 \\ 
      69 & 1.15690 & 1.14556 & 1.13506 & 1.12329 & 1.09494 & 1.07498 \\ 
      70 & 1.15041 & 1.13898 & 1.12878 & 1.11612 & 1.08898 & 1.06510 \\ 
      71 & 1.14323 & 1.13212 & 1.12200 & 1.10930 & 1.08173 & 1.05237 \\ 
      72 & 1.13657 & 1.12558 & 1.11555 & 1.10359 & 1.07485 & 1.05112 \\ 
      73 & 1.12986 & 1.11878 & 1.10884 & 1.09758 & 1.06943 & 1.04447 \\ 
      74 & 1.12381 & 1.11304 & 1.10299 & 1.09112 & 1.06410 & 1.03853 \\ 
      75 & 1.11769 & 1.10702 & 1.09707 & 1.08533 & 1.05778 & 1.04000 \\ 
      76 & 1.11165 & 1.10059 & 1.09107 & 1.07877 & 1.05315 & 1.03116 \\ 
      77 & 1.10525 & 1.09462 & 1.08501 & 1.07320 & 1.04638 & 1.02749 \\ 
      78 & 1.16650 & 1.15631 & 1.14678 & 1.13560 & 1.11012 & 1.08695 \\ 
      79 & 1.16055 & 1.15014 & 1.14084 & 1.12931 & 1.10383 & 1.08348 \\ 
      80 & 1.15453 & 1.14437 & 1.13484 & 1.12344 & 1.09922 & 1.07344 \\ 
      81 & 1.14906 & 1.13900 & 1.12986 & 1.11888 & 1.09282 & 1.07194 \\ 
      82 & 1.14307 & 1.13311 & 1.12433 & 1.11288 & 1.08864 & 1.06856 \\ 
      83 & 1.13805 & 1.12818 & 1.11932 & 1.10814 & 1.08434 & 1.06474 \\ 
      84 & 1.13265 & 1.12259 & 1.11338 & 1.10247 & 1.07851 & 1.05782 \\ 
      85 & 1.12720 & 1.11696 & 1.10754 & 1.09689 & 1.07170 & 1.05190 \\ 
      86 & 1.12196 & 1.11195 & 1.10289 & 1.09248 & 1.06963 & 1.04070 \\ 
      87 & 1.11692 & 1.10715 & 1.09856 & 1.08733 & 1.06223 & 1.04082 \\ 
      88 & 1.11170 & 1.10176 & 1.09310 & 1.08226 & 1.05850 & 1.03719 \\ 
      89 & 1.10668 & 1.09670 & 1.08787 & 1.07650 & 1.05328 & 1.03055 \\ 
      90 & 1.10198 & 1.09235 & 1.08346 & 1.07284 & 1.04901 & 1.02914 \\ 
      91 & 1.09733 & 1.08767 & 1.07910 & 1.06883 & 1.04516 & 1.02210 \\ 
      92 & 1.15064 & 1.14130 & 1.13268 & 1.12228 & 1.09983 & 1.07999 \\ 
      93 & 1.14591 & 1.13666 & 1.12880 & 1.11840 & 1.09481 & 1.07619 \\ 
      94 & 1.14113 & 1.13207 & 1.12381 & 1.11374 & 1.09110 & 1.07209 \\ 
      95 & 1.13651 & 1.12720 & 1.11889 & 1.10903 & 1.08632 & 1.06825 \\ 
      96 & 1.13184 & 1.12250 & 1.11404 & 1.10406 & 1.08203 & 1.06272 \\ 
      97 & 1.12754 & 1.11829 & 1.11021 & 1.09980 & 1.07716 & 1.05707 \\ 
      98 & 1.12349 & 1.11454 & 1.10641 & 1.09652 & 1.07434 & 1.05269 \\ 
      99 & 1.11887 & 1.10978 & 1.10183 & 1.09193 & 1.07091 & 1.05387 \\ 
     100 & 1.11470 & 1.10580 & 1.09740 & 1.08760 & 1.06480 & 1.04480 \\ 
     101 & 1.11229 & 1.10332 & 1.09536 & 1.08528 & 1.06346 & 1.04473 \\ 
     102 & 1.10803 & 1.09910 & 1.09122 & 1.08117 & 1.05942 & 1.04081 \\ 
     103 & 1.10377 & 1.09488 & 1.08707 & 1.07706 & 1.05538 & 1.03689 \\ 
     104 & 1.09951 & 1.09066 & 1.08292 & 1.07295 & 1.05134 & 1.03298 \\ 
     105 & 1.09525 & 1.08644 & 1.07878 & 1.06884 & 1.04730 & 1.02906 \\ 
     106 & 1.09099 & 1.08222 & 1.07463 & 1.06473 & 1.04326 & 1.02514 \\ 
     107 & 1.08673 & 1.07800 & 1.07048 & 1.06062 & 1.03922 & 1.02122 \\ 
     108 & 1.13400 & 1.12543 & 1.11806 & 1.10897 & 1.08813 & 1.07073 \\ 
     109 & 1.13053 & 1.12200 & 1.11464 & 1.10560 & 1.08484 & 1.06766 \\ 
     110 & 1.12706 & 1.11857 & 1.11123 & 1.10223 & 1.08154 & 1.06459 \\ 
     111 & 1.12358 & 1.11514 & 1.10782 & 1.09886 & 1.07824 & 1.06152 \\ 
     112 & 1.12011 & 1.11171 & 1.10441 & 1.09550 & 1.07495 & 1.05845 \\ 
     113 & 1.11664 & 1.10828 & 1.10100 & 1.09213 & 1.07165 & 1.05537 \\ 
     114 & 1.11316 & 1.10485 & 1.09759 & 1.08876 & 1.06835 & 1.05230 \\ 
     115 & 1.10969 & 1.10143 & 1.09418 & 1.08539 & 1.06505 & 1.04923 \\ 
     116 & 1.10622 & 1.09800 & 1.09077 & 1.08203 & 1.06176 & 1.04616 \\ 
     117 & 1.10274 & 1.09457 & 1.08735 & 1.07866 & 1.05846 & 1.04309 \\ 
     118 & 1.09927 & 1.09114 & 1.08394 & 1.07529 & 1.05516 & 1.04002 \\ 
     119 & 1.09580 & 1.08771 & 1.08053 & 1.07192 & 1.05187 & 1.03695 \\ 
     120 & 1.09233 & 1.08428 & 1.07712 & 1.06856 & 1.04857 & 1.03388 \\ 
     121 & 1.08885 & 1.08085 & 1.07371 & 1.06519 & 1.04527 & 1.03081 \\ 
     122 & 1.08538 & 1.07742 & 1.07030 & 1.06182 & 1.04197 & 1.02773 \\ 
     123 & 1.08191 & 1.07399 & 1.06689 & 1.05845 & 1.03868 & 1.02466 \\ 
     124 & 1.07843 & 1.07056 & 1.06348 & 1.05509 & 1.03538 & 1.02159 \\ 
     125 & 1.12051 & 1.11296 & 1.10630 & 1.09811 & 1.07776 & 1.05946 \\ 
     126 & 1.11767 & 1.11013 & 1.10349 & 1.09531 & 1.07514 & 1.05713 \\ 
     127 & 1.11482 & 1.10731 & 1.10068 & 1.09251 & 1.07252 & 1.05480 \\ 
     128 & 1.11198 & 1.10448 & 1.09786 & 1.08971 & 1.06990 & 1.05247 \\ 
     129 & 1.10913 & 1.10166 & 1.09505 & 1.08690 & 1.06728 & 1.05014 \\ 
     130 & 1.10628 & 1.09883 & 1.09223 & 1.08410 & 1.06466 & 1.04782 \\ 
     131 & 1.10344 & 1.09601 & 1.08942 & 1.08130 & 1.06204 & 1.04549 \\ 
     132 & 1.10059 & 1.09318 & 1.08661 & 1.07850 & 1.05942 & 1.04316 \\ 
     133 & 1.09775 & 1.09036 & 1.08379 & 1.07570 & 1.05681 & 1.04083 \\ 
     134 & 1.09490 & 1.08753 & 1.08098 & 1.07290 & 1.05419 & 1.03851 \\ 
     135 & 1.09206 & 1.08471 & 1.07816 & 1.07009 & 1.05157 & 1.03618 \\ 
     136 & 1.08921 & 1.08188 & 1.07535 & 1.06729 & 1.04895 & 1.03385 \\ 
     137 & 1.08637 & 1.07906 & 1.07254 & 1.06449 & 1.04633 & 1.03152 \\ 
     138 & 1.08352 & 1.07623 & 1.06972 & 1.06169 & 1.04371 & 1.02919 \\ 
     139 & 1.08068 & 1.07341 & 1.06691 & 1.05889 & 1.04109 & 1.02687 \\ 
     140 & 1.07783 & 1.07058 & 1.06409 & 1.05608 & 1.03847 & 1.02454 \\ 
     141 & 1.07499 & 1.06776 & 1.06128 & 1.05328 & 1.03585 & 1.02221 \\ 
     142 & 1.07214 & 1.06493 & 1.05846 & 1.05048 & 1.03323 & 1.01988 \\ 
     143 & 1.06929 & 1.06211 & 1.05565 & 1.04768 & 1.03061 & 1.01756 \\ 
     144 & 1.10745 & 1.10050 & 1.09418 & 1.08656 & 1.06964 & 1.05314 \\ 
     145 & 1.10505 & 1.09812 & 1.09183 & 1.08422 & 1.06736 & 1.05108 \\ 
     146 & 1.10266 & 1.09574 & 1.08948 & 1.08188 & 1.06508 & 1.04901 \\ 
     147 & 1.10026 & 1.09336 & 1.08712 & 1.07954 & 1.06281 & 1.04694 \\ 
     148 & 1.09787 & 1.09098 & 1.08477 & 1.07720 & 1.06053 & 1.04487 \\ 
     149 & 1.09548 & 1.08860 & 1.08242 & 1.07486 & 1.05825 & 1.04281 \\ 
     150 & 1.09308 & 1.08622 & 1.08006 & 1.07252 & 1.05598 & 1.04074 \\ 
     151 & 1.09069 & 1.08384 & 1.07771 & 1.07018 & 1.05370 & 1.03867 \\ 
     152 & 1.08830 & 1.08146 & 1.07536 & 1.06784 & 1.05142 & 1.03660 \\ 
     153 & 1.08590 & 1.07908 & 1.07300 & 1.06550 & 1.04915 & 1.03454 \\ 
     154 & 1.08351 & 1.07670 & 1.07065 & 1.06316 & 1.04687 & 1.03247 \\ 
     155 & 1.08111 & 1.07432 & 1.06830 & 1.06081 & 1.04459 & 1.03040 \\ 
     156 & 1.07872 & 1.07193 & 1.06594 & 1.05847 & 1.04232 & 1.02833 \\ 
     157 & 1.07633 & 1.06955 & 1.06359 & 1.05613 & 1.04004 & 1.02627 \\ 
     158 & 1.07393 & 1.06717 & 1.06123 & 1.05379 & 1.03776 & 1.02420 \\ 
     159 & 1.07154 & 1.06479 & 1.05888 & 1.05145 & 1.03548 & 1.02213 \\ 
     160 & 1.06915 & 1.06241 & 1.05653 & 1.04911 & 1.03321 & 1.02006 \\ 
     161 & 1.06675 & 1.06003 & 1.05417 & 1.04677 & 1.03093 & 1.01800 \\ 
     162 & 1.06436 & 1.05765 & 1.05182 & 1.04443 & 1.02865 & 1.01593 \\ 
     163 & 1.09895 & 1.09248 & 1.08657 & 1.07949 & 1.06421 & 1.05123 \\ 
     164 & 1.09692 & 1.09046 & 1.08456 & 1.07749 & 1.06225 & 1.04937 \\ 
     165 & 1.09488 & 1.08844 & 1.08255 & 1.07549 & 1.06030 & 1.04752 \\ 
     166 & 1.09285 & 1.08643 & 1.08054 & 1.07349 & 1.05834 & 1.04566 \\ 
     167 & 1.09081 & 1.08441 & 1.07853 & 1.07149 & 1.05638 & 1.04381 \\ 
     168 & 1.08878 & 1.08239 & 1.07652 & 1.06949 & 1.05443 & 1.04196 \\ 
     169 & 1.08674 & 1.08037 & 1.07451 & 1.06749 & 1.05247 & 1.04010 \\ 
     170 & 1.08471 & 1.07836 & 1.07250 & 1.06549 & 1.05052 & 1.03825 \\ 
     171 & 1.08267 & 1.07634 & 1.07049 & 1.06348 & 1.04856 & 1.03640 \\ 
     172 & 1.08064 & 1.07432 & 1.06848 & 1.06148 & 1.04660 & 1.03454 \\ 
     173 & 1.07860 & 1.07231 & 1.06647 & 1.05948 & 1.04465 & 1.03269 \\ 
     174 & 1.07657 & 1.07029 & 1.06446 & 1.05748 & 1.04269 & 1.03084 \\ 
     175 & 1.07453 & 1.06827 & 1.06245 & 1.05548 & 1.04073 & 1.02898 \\ 
     176 & 1.07250 & 1.06626 & 1.06044 & 1.05348 & 1.03878 & 1.02713 \\ 
     177 & 1.07047 & 1.06424 & 1.05843 & 1.05148 & 1.03682 & 1.02528 \\ 
     178 & 1.06843 & 1.06222 & 1.05642 & 1.04948 & 1.03486 & 1.02342 \\ 
     179 & 1.06640 & 1.06020 & 1.05441 & 1.04748 & 1.03291 & 1.02157 \\ 
     180 & 1.06436 & 1.05819 & 1.05240 & 1.04548 & 1.03095 & 1.01971 \\ 
     181 & 1.06233 & 1.05617 & 1.05039 & 1.04348 & 1.02900 & 1.01786 \\ 
     182 & 1.06029 & 1.05415 & 1.04838 & 1.04148 & 1.02704 & 1.01601 \\ 
     183 & 1.05826 & 1.05214 & 1.04637 & 1.03948 & 1.02508 & 1.01415 \\ 
     184 & 1.08994 & 1.08388 & 1.07872 & 1.07242 & 1.05801 & 1.04182 \\ 
     185 & 1.08820 & 1.08216 & 1.07699 & 1.07071 & 1.05634 & 1.04025 \\ 
     186 & 1.08647 & 1.08044 & 1.07526 & 1.06899 & 1.05468 & 1.03868 \\ 
     187 & 1.08473 & 1.07872 & 1.07353 & 1.06728 & 1.05301 & 1.03711 \\ 
     188 & 1.08300 & 1.07700 & 1.07181 & 1.06556 & 1.05134 & 1.03554 \\ 
     189 & 1.08126 & 1.07527 & 1.07008 & 1.06384 & 1.04968 & 1.03397 \\ 
     190 & 1.07952 & 1.07355 & 1.06835 & 1.06213 & 1.04801 & 1.03240 \\ 
     191 & 1.07779 & 1.07183 & 1.06663 & 1.06041 & 1.04635 & 1.03083 \\ 
     192 & 1.07605 & 1.07011 & 1.06490 & 1.05869 & 1.04468 & 1.02926 \\ 
     193 & 1.07432 & 1.06839 & 1.06317 & 1.05698 & 1.04301 & 1.02769 \\ 
     194 & 1.07258 & 1.06666 & 1.06145 & 1.05526 & 1.04135 & 1.02612 \\ 
     195 & 1.07084 & 1.06494 & 1.05972 & 1.05354 & 1.03968 & 1.02455 \\ 
     196 & 1.06911 & 1.06322 & 1.05799 & 1.05183 & 1.03801 & 1.02298 \\ 
     197 & 1.06737 & 1.06150 & 1.05626 & 1.05011 & 1.03635 & 1.02141 \\ 
     198 & 1.06564 & 1.05977 & 1.05454 & 1.04840 & 1.03468 & 1.01984 \\ 
     199 & 1.06390 & 1.05805 & 1.05281 & 1.04668 & 1.03301 & 1.01827 \\ 
     200 & 1.06216 & 1.05633 & 1.05108 & 1.04496 & 1.03135 & 1.01670 \\ 
     201 & 1.06043 & 1.05461 & 1.04936 & 1.04325 & 1.02968 & 1.01513 \\ 
     202 & 1.05869 & 1.05289 & 1.04763 & 1.04153 & 1.02802 & 1.01356 \\ 
     203 & 1.05696 & 1.05116 & 1.04590 & 1.03981 & 1.02635 & 1.01199 \\ 
     204 & 1.05522 & 1.04944 & 1.04417 & 1.03810 & 1.02468 & 1.01042 \\ 
     205 & 1.05348 & 1.04772 & 1.04245 & 1.03638 & 1.02302 & 1.00885 \\ 
     206 & 1.05175 & 1.04600 & 1.04072 & 1.03466 & 1.02135 & 1.00728 \\ 
     207 & 1.08098 & 1.07536 & 1.07029 & 1.06418 & 1.05092 & 1.03650 \\ 
     208 & 1.07947 & 1.07385 & 1.06879 & 1.06268 & 1.04946 & 1.03508 \\ 
     209 & 1.07795 & 1.07234 & 1.06729 & 1.06119 & 1.04800 & 1.03365 \\ 
     210 & 1.07643 & 1.07083 & 1.06579 & 1.05969 & 1.04654 & 1.03223 \\ 
     211 & 1.07492 & 1.06932 & 1.06429 & 1.05820 & 1.04508 & 1.03081 \\ 
     212 & 1.07340 & 1.06781 & 1.06279 & 1.05670 & 1.04361 & 1.02938 \\ 
     213 & 1.07189 & 1.06630 & 1.06129 & 1.05521 & 1.04215 & 1.02796 \\ 
     214 & 1.07037 & 1.06479 & 1.05979 & 1.05371 & 1.04069 & 1.02654 \\ 
     215 & 1.06886 & 1.06328 & 1.05829 & 1.05222 & 1.03923 & 1.02511 \\ 
     216 & 1.06734 & 1.06177 & 1.05679 & 1.05073 & 1.03777 & 1.02369 \\ 
     217 & 1.06582 & 1.06026 & 1.05529 & 1.04923 & 1.03631 & 1.02227 \\ 
     218 & 1.06431 & 1.05875 & 1.05379 & 1.04774 & 1.03484 & 1.02084 \\ 
     219 & 1.06279 & 1.05724 & 1.05229 & 1.04624 & 1.03338 & 1.01942 \\ 
     220 & 1.06128 & 1.05573 & 1.05078 & 1.04475 & 1.03192 & 1.01800 \\ 
     221 & 1.05976 & 1.05422 & 1.04928 & 1.04325 & 1.03046 & 1.01657 \\ 
     222 & 1.05825 & 1.05271 & 1.04778 & 1.04176 & 1.02900 & 1.01515 \\ 
     223 & 1.05673 & 1.05120 & 1.04628 & 1.04026 & 1.02753 & 1.01373 \\ 
     224 & 1.05521 & 1.04969 & 1.04478 & 1.03877 & 1.02607 & 1.01230 \\ 
     225 & 1.05370 & 1.04818 & 1.04328 & 1.03727 & 1.02461 & 1.01088 \\ 
     226 & 1.05218 & 1.04667 & 1.04178 & 1.03578 & 1.02315 & 1.00946 \\ 
     227 & 1.05067 & 1.04516 & 1.04028 & 1.03428 & 1.02169 & 1.00803 \\ 
     228 & 1.04915 & 1.04365 & 1.03878 & 1.03279 & 1.02023 & 1.00661 \\ 
     229 & 1.04763 & 1.04214 & 1.03728 & 1.03129 & 1.01876 & 1.00519 \\ 
     230 & 1.07488 & 1.06968 & 1.06493 & 1.05941 & 1.04715 & 1.03673 \\ 
     231 & 1.07355 & 1.06836 & 1.06362 & 1.05811 & 1.04591 & 1.03557 \\ 
     232 & 1.07222 & 1.06704 & 1.06231 & 1.05680 & 1.04466 & 1.03441 \\ 
     233 & 1.07090 & 1.06571 & 1.06100 & 1.05550 & 1.04342 & 1.03325 \\ 
     234 & 1.06957 & 1.06439 & 1.05969 & 1.05419 & 1.04218 & 1.03210 \\ 
     235 & 1.06824 & 1.06307 & 1.05838 & 1.05289 & 1.04094 & 1.03094 \\ 
     236 & 1.06691 & 1.06175 & 1.05707 & 1.05158 & 1.03970 & 1.02978 \\ 
     237 & 1.06559 & 1.06043 & 1.05576 & 1.05028 & 1.03846 & 1.02862 \\ 
     238 & 1.06426 & 1.05911 & 1.05444 & 1.04897 & 1.03722 & 1.02746 \\ 
     239 & 1.06293 & 1.05778 & 1.05313 & 1.04767 & 1.03598 & 1.02630 \\ 
     240 & 1.06161 & 1.05646 & 1.05182 & 1.04636 & 1.03474 & 1.02515 \\ 
     241 & 1.06028 & 1.05514 & 1.05051 & 1.04506 & 1.03350 & 1.02399 \\ 
     242 & 1.05895 & 1.05382 & 1.04920 & 1.04375 & 1.03226 & 1.02283 \\ 
     243 & 1.05763 & 1.05250 & 1.04789 & 1.04244 & 1.03102 & 1.02167 \\ 
     244 & 1.05630 & 1.05118 & 1.04658 & 1.04114 & 1.02978 & 1.02051 \\ 
     245 & 1.05497 & 1.04985 & 1.04527 & 1.03983 & 1.02854 & 1.01935 \\ 
     246 & 1.05364 & 1.04853 & 1.04395 & 1.03853 & 1.02730 & 1.01820 \\ 
     247 & 1.05232 & 1.04721 & 1.04264 & 1.03722 & 1.02606 & 1.01704 \\ 
     248 & 1.05099 & 1.04589 & 1.04133 & 1.03592 & 1.02482 & 1.01588 \\ 
     249 & 1.04966 & 1.04457 & 1.04002 & 1.03461 & 1.02358 & 1.01472 \\ 
     250 & 1.04834 & 1.04325 & 1.03871 & 1.03331 & 1.02234 & 1.01356 \\ 
     251 & 1.04701 & 1.04192 & 1.03740 & 1.03200 & 1.02110 & 1.01240 \\ 
     252 & 1.04568 & 1.04060 & 1.03609 & 1.03070 & 1.01986 & 1.01124 \\ 
     253 & 1.04436 & 1.03928 & 1.03478 & 1.02939 & 1.01862 & 1.01009 \\ 
     254 & 1.04303 & 1.03796 & 1.03346 & 1.02809 & 1.01738 & 1.00893 \\ 
     255 & 1.06817 & 1.06341 & 1.05913 & 1.05390 & 1.04269 & 1.03194 \\ 
     256 & 1.06702 & 1.06225 & 1.05796 & 1.05274 & 1.04155 & 1.03085 \\ 
     257 & 1.06587 & 1.06109 & 1.05680 & 1.05159 & 1.04041 & 1.02975 \\ 
     258 & 1.06471 & 1.05993 & 1.05563 & 1.05043 & 1.03926 & 1.02865 \\ 
     259 & 1.06356 & 1.05877 & 1.05447 & 1.04927 & 1.03812 & 1.02756 \\ 
     260 & 1.06241 & 1.05761 & 1.05330 & 1.04811 & 1.03698 & 1.02646 \\ 
     261 & 1.06126 & 1.05645 & 1.05213 & 1.04695 & 1.03583 & 1.02537 \\ 
     262 & 1.06010 & 1.05529 & 1.05097 & 1.04580 & 1.03469 & 1.02427 \\ 
     263 & 1.05895 & 1.05413 & 1.04980 & 1.04464 & 1.03355 & 1.02318 \\ 
     264 & 1.05780 & 1.05297 & 1.04863 & 1.04348 & 1.03241 & 1.02208 \\ 
     265 & 1.05664 & 1.05181 & 1.04747 & 1.04232 & 1.03126 & 1.02099 \\ 
     266 & 1.05549 & 1.05066 & 1.04630 & 1.04116 & 1.03012 & 1.01989 \\ 
     267 & 1.05434 & 1.04950 & 1.04514 & 1.04000 & 1.02898 & 1.01880 \\ 
     268 & 1.05319 & 1.04834 & 1.04397 & 1.03885 & 1.02783 & 1.01770 \\ 
     269 & 1.05203 & 1.04718 & 1.04280 & 1.03769 & 1.02669 & 1.01660 \\ 
     270 & 1.05088 & 1.04602 & 1.04164 & 1.03653 & 1.02555 & 1.01551 \\ 
     271 & 1.04973 & 1.04486 & 1.04047 & 1.03537 & 1.02441 & 1.01441 \\ 
     272 & 1.04857 & 1.04370 & 1.03930 & 1.03421 & 1.02326 & 1.01332 \\ 
     273 & 1.04742 & 1.04254 & 1.03814 & 1.03306 & 1.02212 & 1.01222 \\ 
     274 & 1.04627 & 1.04138 & 1.03697 & 1.03190 & 1.02098 & 1.01113 \\ 
     275 & 1.04511 & 1.04022 & 1.03581 & 1.03074 & 1.01983 & 1.01003 \\ 
     276 & 1.04396 & 1.03906 & 1.03464 & 1.02958 & 1.01869 & 1.00894 \\ 
     277 & 1.04281 & 1.03790 & 1.03347 & 1.02842 & 1.01755 & 1.00784 \\ 
     278 & 1.04166 & 1.03674 & 1.03231 & 1.02727 & 1.01641 & 1.00674 \\ 
     279 & 1.04050 & 1.03559 & 1.03114 & 1.02611 & 1.01526 & 1.00565 \\ 
     280 & 1.03935 & 1.03443 & 1.02997 & 1.02495 & 1.01412 & 1.00455 \\ 
     281 & 1.06263 & 1.05794 & 1.05390 & 1.04900 & 1.03793 & 1.02781 \\ 
     282 & 1.06161 & 1.05693 & 1.05288 & 1.04798 & 1.03696 & 1.02694 \\ 
     283 & 1.06059 & 1.05591 & 1.05187 & 1.04696 & 1.03600 & 1.02607 \\ 
     284 & 1.05957 & 1.05490 & 1.05085 & 1.04594 & 1.03503 & 1.02520 \\ 
     285 & 1.05855 & 1.05389 & 1.04983 & 1.04492 & 1.03406 & 1.02433 \\ 
     286 & 1.05753 & 1.05288 & 1.04882 & 1.04390 & 1.03310 & 1.02347 \\ 
     287 & 1.05652 & 1.05186 & 1.04780 & 1.04288 & 1.03213 & 1.02260 \\ 
     288 & 1.05550 & 1.05085 & 1.04679 & 1.04186 & 1.03117 & 1.02173 \\ 
     289 & 1.05448 & 1.04984 & 1.04577 & 1.04084 & 1.03020 & 1.02086 \\ 
     290 & 1.05346 & 1.04883 & 1.04475 & 1.03982 & 1.02923 & 1.01999 \\ 
     291 & 1.05244 & 1.04781 & 1.04374 & 1.03880 & 1.02827 & 1.01912 \\ 
     292 & 1.05143 & 1.04680 & 1.04272 & 1.03778 & 1.02730 & 1.01825 \\ 
     293 & 1.05041 & 1.04579 & 1.04170 & 1.03676 & 1.02633 & 1.01738 \\ 
     294 & 1.04939 & 1.04478 & 1.04069 & 1.03574 & 1.02537 & 1.01651 \\ 
     295 & 1.04837 & 1.04376 & 1.03967 & 1.03472 & 1.02440 & 1.01564 \\ 
     296 & 1.04735 & 1.04275 & 1.03865 & 1.03369 & 1.02343 & 1.01477 \\ 
     297 & 1.04633 & 1.04174 & 1.03764 & 1.03267 & 1.02247 & 1.01390 \\ 
     298 & 1.04532 & 1.04073 & 1.03662 & 1.03165 & 1.02150 & 1.01303 \\ 
     299 & 1.04430 & 1.03971 & 1.03561 & 1.03063 & 1.02054 & 1.01217 \\ 
     300 & 1.04328 & 1.03870 & 1.03459 & 1.02961 & 1.01957 & 1.01130 \\ 
     301 & 1.04226 & 1.03769 & 1.03357 & 1.02859 & 1.01860 & 1.01043 \\ 
     302 & 1.04124 & 1.03668 & 1.03256 & 1.02757 & 1.01764 & 1.00956 \\ 
     303 & 1.04023 & 1.03566 & 1.03154 & 1.02655 & 1.01667 & 1.00869 \\ 
     304 & 1.03921 & 1.03465 & 1.03052 & 1.02553 & 1.01570 & 1.00782 \\ 
     305 & 1.03819 & 1.03364 & 1.02951 & 1.02451 & 1.01474 & 1.00695 \\ 
     306 & 1.03717 & 1.03263 & 1.02849 & 1.02349 & 1.01377 & 1.00608 \\ 
     307 & 1.03615 & 1.03161 & 1.02748 & 1.02247 & 1.01280 & 1.00521 \\ 
     308 & 1.03513 & 1.03060 & 1.02646 & 1.02145 & 1.01184 & 1.00434 \\ 
     309 & 1.05700 & 1.05279 & 1.04895 & 1.04417 & 1.03414 & 1.02581 \\ 
     310 & 1.05608 & 1.05187 & 1.04804 & 1.04325 & 1.03324 & 1.02492 \\ 
     311 & 1.05517 & 1.05096 & 1.04712 & 1.04234 & 1.03233 & 1.02402 \\ 
     312 & 1.05426 & 1.05004 & 1.04621 & 1.04143 & 1.03143 & 1.02313 \\ 
     313 & 1.05335 & 1.04913 & 1.04529 & 1.04052 & 1.03052 & 1.02224 \\ 
     314 & 1.05244 & 1.04821 & 1.04438 & 1.03960 & 1.02962 & 1.02135 \\ 
     315 & 1.05153 & 1.04730 & 1.04347 & 1.03869 & 1.02871 & 1.02046 \\ 
     316 & 1.05062 & 1.04638 & 1.04255 & 1.03778 & 1.02781 & 1.01957 \\ 
     317 & 1.04971 & 1.04547 & 1.04164 & 1.03687 & 1.02691 & 1.01868 \\ 
     318 & 1.04880 & 1.04456 & 1.04072 & 1.03595 & 1.02600 & 1.01779 \\ 
     319 & 1.04789 & 1.04364 & 1.03981 & 1.03504 & 1.02510 & 1.01690 \\ 
     320 & 1.04698 & 1.04273 & 1.03889 & 1.03413 & 1.02419 & 1.01601 \\ 
     321 & 1.04606 & 1.04181 & 1.03798 & 1.03322 & 1.02329 & 1.01512 \\ 
     322 & 1.04515 & 1.04090 & 1.03706 & 1.03230 & 1.02238 & 1.01423 \\ 
     323 & 1.04424 & 1.03998 & 1.03615 & 1.03139 & 1.02148 & 1.01333 \\ 
     324 & 1.04333 & 1.03907 & 1.03523 & 1.03048 & 1.02057 & 1.01244 \\ 
     325 & 1.04242 & 1.03816 & 1.03432 & 1.02957 & 1.01967 & 1.01155 \\ 
     326 & 1.04151 & 1.03724 & 1.03341 & 1.02866 & 1.01876 & 1.01066 \\ 
     327 & 1.04060 & 1.03633 & 1.03249 & 1.02774 & 1.01786 & 1.00977 \\ 
     328 & 1.03969 & 1.03541 & 1.03158 & 1.02683 & 1.01695 & 1.00888 \\ 
     329 & 1.03878 & 1.03450 & 1.03066 & 1.02592 & 1.01605 & 1.00799 \\ 
     330 & 1.03787 & 1.03358 & 1.02975 & 1.02501 & 1.01514 & 1.00710 \\ 
     331 & 1.03696 & 1.03267 & 1.02883 & 1.02409 & 1.01424 & 1.00621 \\ 
     332 & 1.03605 & 1.03175 & 1.02792 & 1.02318 & 1.01334 & 1.00532 \\ 
     333 & 1.03513 & 1.03084 & 1.02700 & 1.02227 & 1.01243 & 1.00443 \\ 
     334 & 1.03422 & 1.02993 & 1.02609 & 1.02136 & 1.01153 & 1.00354 \\ 
     335 & 1.03331 & 1.02901 & 1.02518 & 1.02044 & 1.01062 & 1.00264 \\ 
     336 & 1.03240 & 1.02810 & 1.02426 & 1.01953 & 1.00972 & 1.00175 \\ 
     337 & 1.05310 & 1.04897 & 1.04528 & 1.04088 & 1.03216 & 1.02235 \\ 
     338 & 1.05227 & 1.04815 & 1.04446 & 1.04007 & 1.03131 & 1.02156 \\ 
     339 & 1.05145 & 1.04733 & 1.04364 & 1.03926 & 1.03046 & 1.02077 \\ 
     340 & 1.05063 & 1.04651 & 1.04282 & 1.03844 & 1.02961 & 1.01999 \\ 
     341 & 1.04980 & 1.04569 & 1.04200 & 1.03763 & 1.02876 & 1.01920 \\ 
     342 & 1.04898 & 1.04487 & 1.04118 & 1.03682 & 1.02791 & 1.01841 \\ 
     343 & 1.04816 & 1.04405 & 1.04037 & 1.03601 & 1.02706 & 1.01763 \\ 
     344 & 1.04734 & 1.04323 & 1.03955 & 1.03519 & 1.02622 & 1.01684 \\ 
     345 & 1.04651 & 1.04241 & 1.03873 & 1.03438 & 1.02537 & 1.01605 \\ 
     346 & 1.04569 & 1.04159 & 1.03791 & 1.03357 & 1.02452 & 1.01526 \\ 
     347 & 1.04487 & 1.04077 & 1.03709 & 1.03275 & 1.02367 & 1.01448 \\ 
     348 & 1.04404 & 1.03995 & 1.03627 & 1.03194 & 1.02282 & 1.01369 \\ 
     349 & 1.04322 & 1.03913 & 1.03546 & 1.03113 & 1.02197 & 1.01290 \\ 
     350 & 1.04240 & 1.03830 & 1.03464 & 1.03032 & 1.02112 & 1.01212 \\ 
     351 & 1.04158 & 1.03748 & 1.03382 & 1.02950 & 1.02028 & 1.01133 \\ 
     352 & 1.04075 & 1.03666 & 1.03300 & 1.02869 & 1.01943 & 1.01054 \\ 
     353 & 1.03993 & 1.03584 & 1.03218 & 1.02788 & 1.01858 & 1.00975 \\ 
     354 & 1.03911 & 1.03502 & 1.03136 & 1.02707 & 1.01773 & 1.00897 \\ 
     355 & 1.03828 & 1.03420 & 1.03055 & 1.02625 & 1.01688 & 1.00818 \\ 
     356 & 1.03746 & 1.03338 & 1.02973 & 1.02544 & 1.01603 & 1.00739 \\ 
     357 & 1.03664 & 1.03256 & 1.02891 & 1.02463 & 1.01518 & 1.00661 \\ 
     358 & 1.03582 & 1.03174 & 1.02809 & 1.02381 & 1.01434 & 1.00582 \\ 
     359 & 1.03499 & 1.03092 & 1.02727 & 1.02300 & 1.01349 & 1.00503 \\ 
     360 & 1.03417 & 1.03010 & 1.02645 & 1.02219 & 1.01264 & 1.00424 \\ 
     361 & 1.03335 & 1.02928 & 1.02564 & 1.02138 & 1.01179 & 1.00346 \\ 
     362 & 1.03252 & 1.02846 & 1.02482 & 1.02056 & 1.01094 & 1.00267 \\ 
     363 & 1.03170 & 1.02763 & 1.02400 & 1.01975 & 1.01009 & 1.00188 \\ 
     364 & 1.03088 & 1.02681 & 1.02318 & 1.01894 & 1.00924 & 1.00110 \\ 
     365 & 1.03006 & 1.02599 & 1.02236 & 1.01812 & 1.00840 & 1.00031 \\ 
     366 & 1.02923 & 1.02517 & 1.02154 & 1.01731 & 1.00755 & 0.99952 \\ 
     367 & 1.04869 & 1.04481 & 1.04132 & 1.03715 & 1.02815 & 1.01890 \\ 
     368 & 1.04795 & 1.04407 & 1.04059 & 1.03642 & 1.02743 & 1.01826 \\ 
     369 & 1.04722 & 1.04333 & 1.03986 & 1.03569 & 1.02670 & 1.01762 \\ 
     370 & 1.04648 & 1.04260 & 1.03912 & 1.03496 & 1.02598 & 1.01699 \\ 
     371 & 1.04575 & 1.04186 & 1.03839 & 1.03423 & 1.02525 & 1.01635 \\ 
     372 & 1.04501 & 1.04113 & 1.03765 & 1.03350 & 1.02453 & 1.01571 \\ 
     373 & 1.04428 & 1.04039 & 1.03692 & 1.03277 & 1.02380 & 1.01507 \\ 
     374 & 1.04354 & 1.03965 & 1.03619 & 1.03204 & 1.02308 & 1.01443 \\ 
     375 & 1.04281 & 1.03892 & 1.03545 & 1.03131 & 1.02235 & 1.01379 \\ 
     376 & 1.04207 & 1.03818 & 1.03472 & 1.03058 & 1.02162 & 1.01315 \\ 
     377 & 1.04134 & 1.03744 & 1.03398 & 1.02985 & 1.02090 & 1.01251 \\ 
     378 & 1.04060 & 1.03671 & 1.03325 & 1.02912 & 1.02017 & 1.01187 \\ 
     379 & 1.03987 & 1.03597 & 1.03252 & 1.02839 & 1.01945 & 1.01123 \\ 
     380 & 1.03913 & 1.03524 & 1.03178 & 1.02766 & 1.01872 & 1.01059 \\ 
     381 & 1.03840 & 1.03450 & 1.03105 & 1.02693 & 1.01800 & 1.00995 \\ 
     382 & 1.03766 & 1.03376 & 1.03031 & 1.02620 & 1.01727 & 1.00931 \\ 
     383 & 1.03693 & 1.03303 & 1.02958 & 1.02547 & 1.01655 & 1.00868 \\ 
     384 & 1.03619 & 1.03229 & 1.02884 & 1.02474 & 1.01582 & 1.00804 \\ 
     385 & 1.03546 & 1.03156 & 1.02811 & 1.02401 & 1.01510 & 1.00740 \\ 
     386 & 1.03472 & 1.03082 & 1.02738 & 1.02328 & 1.01437 & 1.00676 \\ 
     387 & 1.03399 & 1.03008 & 1.02664 & 1.02255 & 1.01364 & 1.00612 \\ 
     388 & 1.03325 & 1.02935 & 1.02591 & 1.02182 & 1.01292 & 1.00548 \\ 
     389 & 1.03252 & 1.02861 & 1.02517 & 1.02109 & 1.01219 & 1.00484 \\ 
     390 & 1.03178 & 1.02787 & 1.02444 & 1.02036 & 1.01147 & 1.00420 \\ 
     391 & 1.03105 & 1.02714 & 1.02371 & 1.01963 & 1.01074 & 1.00356 \\ 
     392 & 1.03031 & 1.02640 & 1.02297 & 1.01890 & 1.01002 & 1.00292 \\ 
     393 & 1.02958 & 1.02567 & 1.02224 & 1.01817 & 1.00929 & 1.00228 \\ 
     394 & 1.02884 & 1.02493 & 1.02150 & 1.01744 & 1.00857 & 1.00164 \\ 
     395 & 1.02811 & 1.02419 & 1.02077 & 1.01671 & 1.00784 & 1.00101 \\ 
     396 & 1.02737 & 1.02346 & 1.02004 & 1.01598 & 1.00711 & 1.00037 \\ 
     397 & 1.02664 & 1.02272 & 1.01930 & 1.01525 & 1.00639 & 0.99973 \\ 
     398 & 1.04487 & 1.04108 & 1.03781 & 1.03391 & 1.02503 & 1.01735 \\ 
     399 & 1.04421 & 1.04042 & 1.03715 & 1.03325 & 1.02437 & 1.01671 \\ 
     400 & 1.04354 & 1.03976 & 1.03648 & 1.03258 & 1.02371 & 1.01607 \\ 
     401 & 1.04288 & 1.03910 & 1.03582 & 1.03192 & 1.02305 & 1.01543 \\ 
     402 & 1.04221 & 1.03844 & 1.03516 & 1.03126 & 1.02239 & 1.01479 \\ 
     403 & 1.04155 & 1.03778 & 1.03449 & 1.03059 & 1.02173 & 1.01415 \\ 
     404 & 1.04088 & 1.03711 & 1.03383 & 1.02993 & 1.02107 & 1.01351 \\ 
     405 & 1.04022 & 1.03645 & 1.03317 & 1.02926 & 1.02041 & 1.01287 \\ 
     406 & 1.03955 & 1.03579 & 1.03250 & 1.02860 & 1.01975 & 1.01223 \\ 
     407 & 1.03889 & 1.03513 & 1.03184 & 1.02793 & 1.01909 & 1.01159 \\ 
     408 & 1.03822 & 1.03447 & 1.03118 & 1.02727 & 1.01843 & 1.01095 \\ 
     409 & 1.03756 & 1.03380 & 1.03051 & 1.02660 & 1.01777 & 1.01031 \\ 
     410 & 1.03689 & 1.03314 & 1.02985 & 1.02594 & 1.01710 & 1.00967 \\ 
     411 & 1.03623 & 1.03248 & 1.02919 & 1.02528 & 1.01644 & 1.00903 \\ 
     412 & 1.03556 & 1.03182 & 1.02852 & 1.02461 & 1.01578 & 1.00839 \\ 
     413 & 1.03490 & 1.03116 & 1.02786 & 1.02395 & 1.01512 & 1.00775 \\ 
     414 & 1.03423 & 1.03049 & 1.02720 & 1.02328 & 1.01446 & 1.00711 \\ 
     415 & 1.03357 & 1.02983 & 1.02653 & 1.02262 & 1.01380 & 1.00647 \\ 
     416 & 1.03290 & 1.02917 & 1.02587 & 1.02195 & 1.01314 & 1.00583 \\ 
     417 & 1.03224 & 1.02851 & 1.02521 & 1.02129 & 1.01248 & 1.00519 \\ 
     418 & 1.03158 & 1.02785 & 1.02454 & 1.02062 & 1.01182 & 1.00455 \\ 
     419 & 1.03091 & 1.02718 & 1.02388 & 1.01996 & 1.01116 & 1.00391 \\ 
     420 & 1.03025 & 1.02652 & 1.02322 & 1.01929 & 1.01050 & 1.00327 \\ 
     421 & 1.02958 & 1.02586 & 1.02255 & 1.01863 & 1.00984 & 1.00263 \\ 
     422 & 1.02892 & 1.02520 & 1.02189 & 1.01797 & 1.00918 & 1.00199 \\ 
     423 & 1.02825 & 1.02454 & 1.02123 & 1.01730 & 1.00852 & 1.00135 \\ 
     424 & 1.02759 & 1.02387 & 1.02056 & 1.01664 & 1.00786 & 1.00071 \\ 
     425 & 1.02692 & 1.02321 & 1.01990 & 1.01597 & 1.00720 & 1.00007 \\ 
     426 & 1.02626 & 1.02255 & 1.01924 & 1.01531 & 1.00654 & 0.99943 \\ 
     427 & 1.02559 & 1.02189 & 1.01857 & 1.01464 & 1.00588 & 0.99879 \\ 
     428 & 1.02493 & 1.02123 & 1.01791 & 1.01398 & 1.00522 & 0.99815 \\ 
     429 & 1.02426 & 1.02056 & 1.01725 & 1.01331 & 1.00455 & 0.99751 \\ 
     430 & 1.02360 & 1.01990 & 1.01658 & 1.01265 & 1.00389 & 0.99687 \\ 
     431 & 1.04110 & 1.03745 & 1.03418 & 1.03052 & 1.02236 & 1.01457 \\ 
     432 & 1.04050 & 1.03685 & 1.03359 & 1.02991 & 1.02176 & 1.01399 \\ 
     433 & 1.03990 & 1.03625 & 1.03299 & 1.02931 & 1.02116 & 1.01342 \\ 
     434 & 1.03929 & 1.03565 & 1.03239 & 1.02871 & 1.02056 & 1.01285 \\ 
     435 & 1.03869 & 1.03505 & 1.03179 & 1.02810 & 1.01995 & 1.01227 \\ 
     436 & 1.03809 & 1.03445 & 1.03119 & 1.02750 & 1.01935 & 1.01170 \\ 
     437 & 1.03749 & 1.03385 & 1.03059 & 1.02690 & 1.01875 & 1.01112 \\ 
     438 & 1.03688 & 1.03325 & 1.02999 & 1.02630 & 1.01815 & 1.01055 \\ 
     439 & 1.03628 & 1.03265 & 1.02940 & 1.02569 & 1.01755 & 1.00997 \\ 
     440 & 1.03568 & 1.03205 & 1.02880 & 1.02509 & 1.01694 & 1.00940 \\ 
     441 & 1.03507 & 1.03145 & 1.02820 & 1.02449 & 1.01634 & 1.00883 \\ 
     442 & 1.03447 & 1.03085 & 1.02760 & 1.02388 & 1.01574 & 1.00825 \\ 
     443 & 1.03387 & 1.03025 & 1.02700 & 1.02328 & 1.01514 & 1.00768 \\ 
     444 & 1.03327 & 1.02965 & 1.02640 & 1.02268 & 1.01454 & 1.00710 \\ 
     445 & 1.03266 & 1.02905 & 1.02580 & 1.02207 & 1.01393 & 1.00653 \\ 
     446 & 1.03206 & 1.02845 & 1.02521 & 1.02147 & 1.01333 & 1.00596 \\ 
     447 & 1.03146 & 1.02785 & 1.02461 & 1.02087 & 1.01273 & 1.00538 \\ 
     448 & 1.03086 & 1.02725 & 1.02401 & 1.02026 & 1.01213 & 1.00481 \\ 
     449 & 1.03025 & 1.02665 & 1.02341 & 1.01966 & 1.01152 & 1.00423 \\ 
     450 & 1.02965 & 1.02605 & 1.02281 & 1.01906 & 1.01092 & 1.00366 \\ 
     451 & 1.02905 & 1.02545 & 1.02221 & 1.01845 & 1.01032 & 1.00309 \\ 
     452 & 1.02844 & 1.02485 & 1.02161 & 1.01785 & 1.00972 & 1.00251 \\ 
     453 & 1.02784 & 1.02425 & 1.02102 & 1.01725 & 1.00912 & 1.00194 \\ 
     454 & 1.02724 & 1.02365 & 1.02042 & 1.01664 & 1.00851 & 1.00136 \\ 
     455 & 1.02664 & 1.02305 & 1.01982 & 1.01604 & 1.00791 & 1.00079 \\ 
     456 & 1.02603 & 1.02245 & 1.01922 & 1.01544 & 1.00731 & 1.00022 \\ 
     457 & 1.02543 & 1.02185 & 1.01862 & 1.01484 & 1.00671 & 0.99964 \\ 
     458 & 1.02483 & 1.02125 & 1.01802 & 1.01423 & 1.00611 & 0.99907 \\ 
     459 & 1.02423 & 1.02065 & 1.01742 & 1.01363 & 1.00550 & 0.99849 \\ 
     460 & 1.02362 & 1.02005 & 1.01683 & 1.01303 & 1.00490 & 0.99792 \\ 
     461 & 1.02302 & 1.01945 & 1.01623 & 1.01242 & 1.00430 & 0.99735 \\ 
     462 & 1.02242 & 1.01885 & 1.01563 & 1.01182 & 1.00370 & 0.99677 \\ 
     463 & 1.02182 & 1.01825 & 1.01503 & 1.01122 & 1.00310 & 0.99620 \\ 
     464 & 1.02121 & 1.01765 & 1.01443 & 1.01061 & 1.00249 & 0.99562 \\ 
     465 & 1.03769 & 1.03430 & 1.03122 & 1.02757 & 1.01995 & 1.01307 \\ 
     466 & 1.03714 & 1.03375 & 1.03067 & 1.02702 & 1.01939 & 1.01253 \\ 
     467 & 1.03659 & 1.03319 & 1.03012 & 1.02647 & 1.01884 & 1.01199 \\ 
     468 & 1.03604 & 1.03264 & 1.02957 & 1.02592 & 1.01828 & 1.01146 \\ 
     469 & 1.03549 & 1.03209 & 1.02902 & 1.02537 & 1.01773 & 1.01092 \\ 
     470 & 1.03494 & 1.03154 & 1.02847 & 1.02482 & 1.01717 & 1.01038 \\ 
     471 & 1.03439 & 1.03099 & 1.02791 & 1.02427 & 1.01662 & 1.00984 \\ 
     472 & 1.03384 & 1.03043 & 1.02736 & 1.02372 & 1.01606 & 1.00931 \\ 
     473 & 1.03329 & 1.02988 & 1.02681 & 1.02317 & 1.01551 & 1.00877 \\ 
     474 & 1.03274 & 1.02933 & 1.02626 & 1.02262 & 1.01495 & 1.00823 \\ 
     475 & 1.03219 & 1.02878 & 1.02571 & 1.02207 & 1.01440 & 1.00769 \\ 
     476 & 1.03164 & 1.02823 & 1.02516 & 1.02152 & 1.01384 & 1.00716 \\ 
     477 & 1.03109 & 1.02767 & 1.02461 & 1.02097 & 1.01329 & 1.00662 \\ 
     478 & 1.03054 & 1.02712 & 1.02406 & 1.02042 & 1.01273 & 1.00608 \\ 
     479 & 1.02999 & 1.02657 & 1.02351 & 1.01987 & 1.01218 & 1.00554 \\ 
     480 & 1.02944 & 1.02602 & 1.02296 & 1.01932 & 1.01163 & 1.00501 \\ 
     481 & 1.02889 & 1.02547 & 1.02241 & 1.01876 & 1.01107 & 1.00447 \\ 
     482 & 1.02834 & 1.02491 & 1.02186 & 1.01821 & 1.01052 & 1.00393 \\ 
     483 & 1.02779 & 1.02436 & 1.02131 & 1.01766 & 1.00996 & 1.00339 \\ 
     484 & 1.02724 & 1.02381 & 1.02076 & 1.01711 & 1.00941 & 1.00286 \\ 
     485 & 1.02669 & 1.02326 & 1.02021 & 1.01656 & 1.00885 & 1.00232 \\ 
     486 & 1.02614 & 1.02271 & 1.01966 & 1.01601 & 1.00830 & 1.00178 \\ 
     487 & 1.02559 & 1.02215 & 1.01911 & 1.01546 & 1.00774 & 1.00124 \\ 
     488 & 1.02504 & 1.02160 & 1.01856 & 1.01491 & 1.00719 & 1.00071 \\ 
     489 & 1.02449 & 1.02105 & 1.01801 & 1.01436 & 1.00663 & 1.00017 \\ 
     490 & 1.02394 & 1.02050 & 1.01746 & 1.01381 & 1.00608 & 0.99963 \\ 
     491 & 1.02340 & 1.01995 & 1.01691 & 1.01326 & 1.00552 & 0.99910 \\ 
     492 & 1.02285 & 1.01939 & 1.01636 & 1.01271 & 1.00497 & 0.99856 \\ 
     493 & 1.02230 & 1.01884 & 1.01580 & 1.01216 & 1.00441 & 0.99802 \\ 
     494 & 1.02175 & 1.01829 & 1.01525 & 1.01161 & 1.00386 & 0.99748 \\ 
     495 & 1.02120 & 1.01774 & 1.01470 & 1.01106 & 1.00330 & 0.99695 \\ 
     496 & 1.02065 & 1.01719 & 1.01415 & 1.01051 & 1.00275 & 0.99641 \\ 
     497 & 1.02010 & 1.01663 & 1.01360 & 1.00996 & 1.00219 & 0.99587 \\ 
     498 & 1.01955 & 1.01608 & 1.01305 & 1.00941 & 1.00164 & 0.99533 \\ 
     499 & 1.01900 & 1.01553 & 1.01250 & 1.00886 & 1.00108 & 0.99480 \\ 
     500 & 1.03464 & 1.03135 & 1.02835 & 1.02489 & 1.01768 & 1.01220 \\ 
   \hline
\hline
\label{table:s2}
\end{longtable}
\normalsize

\tiny 
\renewcommand*{\arraystretch}{1.22}
\begin{longtable}{ccccccc}
\caption{\textbf{Showing the $p$-th quantiles of the hybrid p-values}} \\

  \hline
Sample Sizes & p=0.1 & p=0.05 & p=0.025 & p=0.01 & p=0.001 & p=0.0001 \\ 
  \hline
  \endhead
  \hline
\multicolumn{7}{r@{}}{continued to next page\ldots}\\
\endfoot    
\hline
\endlastfoot    

			20 & 0.06682 & 0.03219 & 0.01591 & 0.00659 & 0.00065 & 0.00007 \\ 
      21 & 0.06464 & 0.03226 & 0.01573 & 0.00632 & 0.00062 & 0.00005 \\ 
      22 & 0.06512 & 0.03175 & 0.01531 & 0.00620 & 0.00062 & 0.00006 \\ 
      23 & 0.06590 & 0.03182 & 0.01594 & 0.00618 & 0.00070 & 0.00007 \\ 
      24 & 0.06512 & 0.03226 & 0.01601 & 0.00647 & 0.00062 & 0.00006 \\ 
      25 & 0.06479 & 0.03165 & 0.01585 & 0.00622 & 0.00064 & 0.00006 \\ 
      26 & 0.06357 & 0.03102 & 0.01566 & 0.00631 & 0.00063 & 0.00006 \\ 
      27 & 0.06366 & 0.03112 & 0.01531 & 0.00607 & 0.00062 & 0.00007 \\ 
      28 & 0.06309 & 0.03098 & 0.01512 & 0.00599 & 0.00061 & 0.00007 \\ 
      29 & 0.06346 & 0.03038 & 0.01503 & 0.00600 & 0.00060 & 0.00006 \\ 
      30 & 0.06293 & 0.03024 & 0.01495 & 0.00590 & 0.00057 & 0.00005 \\ 
      31 & 0.06257 & 0.03042 & 0.01481 & 0.00586 & 0.00058 & 0.00007 \\ 
      32 & 0.06206 & 0.03057 & 0.01498 & 0.00579 & 0.00058 & 0.00006 \\ 
      33 & 0.06207 & 0.03016 & 0.01502 & 0.00606 & 0.00056 & 0.00005 \\ 
      34 & 0.06169 & 0.03010 & 0.01497 & 0.00592 & 0.00058 & 0.00005 \\ 
      35 & 0.06171 & 0.03003 & 0.01490 & 0.00584 & 0.00059 & 0.00005 \\ 
      36 & 0.06153 & 0.03016 & 0.01465 & 0.00572 & 0.00058 & 0.00007 \\ 
      37 & 0.06146 & 0.02965 & 0.01465 & 0.00574 & 0.00058 & 0.00005 \\ 
      38 & 0.06027 & 0.02944 & 0.01454 & 0.00565 & 0.00057 & 0.00006 \\ 
      39 & 0.06069 & 0.02942 & 0.01447 & 0.00566 & 0.00055 & 0.00005 \\ 
      40 & 0.06032 & 0.02926 & 0.01420 & 0.00564 & 0.00056 & 0.00005 \\ 
      41 & 0.06083 & 0.02960 & 0.01451 & 0.00572 & 0.00052 & 0.00006 \\ 
      42 & 0.05989 & 0.02923 & 0.01446 & 0.00574 & 0.00055 & 0.00005 \\ 
      43 & 0.06029 & 0.02928 & 0.01470 & 0.00577 & 0.00054 & 0.00005 \\ 
      44 & 0.06045 & 0.02904 & 0.01431 & 0.00558 & 0.00055 & 0.00006 \\ 
      45 & 0.05955 & 0.02907 & 0.01417 & 0.00570 & 0.00057 & 0.00005 \\ 
      46 & 0.05950 & 0.02885 & 0.01419 & 0.00553 & 0.00054 & 0.00006 \\ 
      47 & 0.05953 & 0.02891 & 0.01410 & 0.00558 & 0.00055 & 0.00006 \\ 
      48 & 0.05962 & 0.02908 & 0.01415 & 0.00553 & 0.00055 & 0.00005 \\ 
      49 & 0.05952 & 0.02874 & 0.01410 & 0.00549 & 0.00054 & 0.00005 \\ 
      50 & 0.05903 & 0.02858 & 0.01417 & 0.00561 & 0.00053 & 0.00005 \\ 
      51 & 0.05913 & 0.02857 & 0.01412 & 0.00558 & 0.00055 & 0.00006 \\ 
      52 & 0.05933 & 0.02896 & 0.01406 & 0.00540 & 0.00053 & 0.00005 \\ 
      53 & 0.05925 & 0.02884 & 0.01417 & 0.00557 & 0.00053 & 0.00004 \\ 
      54 & 0.05918 & 0.02879 & 0.01408 & 0.00553 & 0.00052 & 0.00005 \\ 
      55 & 0.05867 & 0.02861 & 0.01393 & 0.00550 & 0.00055 & 0.00005 \\ 
      56 & 0.05871 & 0.02862 & 0.01423 & 0.00554 & 0.00052 & 0.00005 \\ 
      57 & 0.05874 & 0.02838 & 0.01382 & 0.00552 & 0.00056 & 0.00005 \\ 
      58 & 0.05874 & 0.02865 & 0.01394 & 0.00543 & 0.00053 & 0.00005 \\ 
      59 & 0.05868 & 0.02845 & 0.01408 & 0.00550 & 0.00052 & 0.00006 \\ 
      60 & 0.05843 & 0.02840 & 0.01378 & 0.00545 & 0.00053 & 0.00005 \\ 
      61 & 0.05849 & 0.02840 & 0.01398 & 0.00547 & 0.00055 & 0.00006 \\ 
      62 & 0.05806 & 0.02831 & 0.01390 & 0.00541 & 0.00052 & 0.00005 \\ 
      63 & 0.05810 & 0.02812 & 0.01372 & 0.00546 & 0.00053 & 0.00005 \\ 
      64 & 0.05832 & 0.02852 & 0.01391 & 0.00544 & 0.00053 & 0.00005 \\ 
      65 & 0.05770 & 0.02831 & 0.01386 & 0.00543 & 0.00054 & 0.00006 \\ 
      66 & 0.05831 & 0.02817 & 0.01378 & 0.00543 & 0.00052 & 0.00005 \\ 
      67 & 0.05779 & 0.02805 & 0.01379 & 0.00539 & 0.00051 & 0.00005 \\ 
      68 & 0.05776 & 0.02800 & 0.01379 & 0.00551 & 0.00054 & 0.00005 \\ 
      69 & 0.05768 & 0.02780 & 0.01356 & 0.00534 & 0.00052 & 0.00005 \\ 
      70 & 0.05776 & 0.02806 & 0.01382 & 0.00536 & 0.00054 & 0.00005 \\ 
      71 & 0.05744 & 0.02778 & 0.01368 & 0.00536 & 0.00053 & 0.00005 \\ 
      72 & 0.05746 & 0.02776 & 0.01363 & 0.00540 & 0.00053 & 0.00005 \\ 
      73 & 0.05736 & 0.02798 & 0.01370 & 0.00544 & 0.00052 & 0.00005 \\ 
      74 & 0.05709 & 0.02770 & 0.01358 & 0.00537 & 0.00053 & 0.00005 \\ 
      75 & 0.05712 & 0.02766 & 0.01352 & 0.00526 & 0.00051 & 0.00005 \\ 
      76 & 0.05675 & 0.02759 & 0.01362 & 0.00538 & 0.00053 & 0.00006 \\ 
      77 & 0.05669 & 0.02769 & 0.01355 & 0.00531 & 0.00053 & 0.00005 \\ 
      78 & 0.05698 & 0.02765 & 0.01361 & 0.00534 & 0.00052 & 0.00005 \\ 
      79 & 0.05656 & 0.02766 & 0.01354 & 0.00538 & 0.00053 & 0.00005 \\ 
      80 & 0.05704 & 0.02791 & 0.01359 & 0.00534 & 0.00053 & 0.00005 \\ 
      81 & 0.05709 & 0.02756 & 0.01352 & 0.00530 & 0.00051 & 0.00005 \\ 
      82 & 0.05706 & 0.02765 & 0.01360 & 0.00530 & 0.00051 & 0.00005 \\ 
      83 & 0.05659 & 0.02730 & 0.01336 & 0.00527 & 0.00052 & 0.00005 \\ 
      84 & 0.05707 & 0.02768 & 0.01363 & 0.00539 & 0.00051 & 0.00005 \\ 
      85 & 0.05680 & 0.02755 & 0.01347 & 0.00526 & 0.00052 & 0.00005 \\ 
      86 & 0.05686 & 0.02750 & 0.01362 & 0.00538 & 0.00054 & 0.00005 \\ 
      87 & 0.05707 & 0.02766 & 0.01352 & 0.00535 & 0.00053 & 0.00006 \\ 
      88 & 0.05655 & 0.02756 & 0.01347 & 0.00530 & 0.00051 & 0.00005 \\ 
      89 & 0.05685 & 0.02762 & 0.01352 & 0.00532 & 0.00051 & 0.00005 \\ 
      90 & 0.05624 & 0.02709 & 0.01332 & 0.00524 & 0.00053 & 0.00005 \\ 
      91 & 0.05628 & 0.02728 & 0.01343 & 0.00527 & 0.00053 & 0.00005 \\ 
      92 & 0.05639 & 0.02727 & 0.01346 & 0.00529 & 0.00051 & 0.00005 \\ 
      93 & 0.05645 & 0.02736 & 0.01344 & 0.00530 & 0.00051 & 0.00005 \\ 
      94 & 0.05645 & 0.02751 & 0.01347 & 0.00536 & 0.00051 & 0.00005 \\ 
      95 & 0.05639 & 0.02755 & 0.01353 & 0.00530 & 0.00051 & 0.00005 \\ 
      96 & 0.05637 & 0.02727 & 0.01341 & 0.00529 & 0.00051 & 0.00005 \\ 
      97 & 0.05646 & 0.02716 & 0.01337 & 0.00528 & 0.00051 & 0.00005 \\ 
      98 & 0.05668 & 0.02731 & 0.01350 & 0.00524 & 0.00050 & 0.00005 \\ 
      99 & 0.05615 & 0.02736 & 0.01346 & 0.00527 & 0.00053 & 0.00006 \\ 
     100 & 0.05616 & 0.02737 & 0.01343 & 0.00527 & 0.00052 & 0.00005 \\ 
   \hline
\hline
\label{table:s3}
\end{longtable}
\normalsize

\bibliography{rankcover}